\documentclass[12pt, a4paper]{article}

\usepackage[T1, T2A]{fontenc}
\usepackage[cp1251]{inputenc}
\usepackage[english]{babel}
\usepackage{amsmath, amssymb, xcolor}
\usepackage[left=3cm, right=1.5cm, top=2cm, bottom=2cm]{geometry}
\usepackage[displaymath]{lineno}
\usepackage{graphicx}

\newcommand\rem[1]{}

\def\ga{\mathrel{\mathchoice {\vcenter{\offinterlineskip\halign{\hfil
$\displaystyle##$\hfil\cr>\cr\sim\cr}}}
{\vcenter{\offinterlineskip\halign{\hfil$\textstyle##$\hfil\cr
>\cr\sim\cr}}}
{\vcenter{\offinterlineskip\halign{\hfil$\scriptstyle##$\hfil\cr
>\cr\sim\cr}}}
{\vcenter{\offinterlineskip\halign{\hfil$\scriptscriptstyle##$\hfil\cr
>\cr\sim\cr}}}}}

\def\utw{\smash{\rlap{\lower5pt\hbox{$\sim$}}}}
\def\udtw{\smash{\rlap{\lower6pt\hbox{$\approx$}}}}

\def\diameter{{\ifmmode\mathchoice
{\ooalign{\hfil\hbox{$\displaystyle/$}\hfil\crcr
{\hbox{$\displaystyle\mathchar"20D$}}}}
{\ooalign{\hfil\hbox{$\textstyle/$}\hfil\crcr
{\hbox{$\textstyle\mathchar"20D$}}}}
{\ooalign{\hfil\hbox{$\scriptstyle/$}\hfil\crcr
{\hbox{$\scriptstyle\mathchar"20D$}}}}
{\ooalign{\hfil\hbox{$\scriptscriptstyle/$}\hfil\crcr
{\hbox{$\scriptscriptstyle\mathchar"20D$}}}}
\else{\ooalign{\hfil/\hfil\crcr\mathhexbox20D}}%
\fi}}

\begin{document}

\begin{center}
{\Large\noindent
Dust Temperature Profiles in Dense Cores Associated With Massive Star-Forming Regions
}

{\it L.\,E. Pirogov}

\small{Institute of Applied Physics of the Russian Academy of Sciences} \\

\small{Published in Russian in Izvestia Vuzov. Radofizika}
\small\rm{(DOI: $10.52452/00213462\_2021\_64\_12\_954$)} \\

\it{Special issue dedicated to the 90th anniversary
of professor Albert G. Kislyakov (1931-2019)}

\end{center}

Using the APEX-12m telescope, continuum maps
at a wavelength of 350~$\mu$m of eight gas-dust clouds from
the southern hemisphere are obtained.
The clouds are associated with the regions of massive star and star cluster
formation and have dense cores.
The core sizes estimated at half of the maximum intensity level 
are $\sim 0.1-0.2$~pc.
The core masses and gas mean densities lie in the ranges:
$\sim 20-1000~M_{\odot}$ and $\sim (0.3-7.3)\times 10^6$~cm$^{-3}$,
respectively.
A comparison of the 350~$\mu$m data with the data
of observations at a wavelength of 1.2~mm has been carried out.
From the intensity ratios at two wavelengths convolved
to the same angular resolution,
the spatial distributions of the average dust temperatures 
on the line of sight are calculated.
Dust temperature maps for most objects
correlate with intensity distributions at 350~$\mu$m.
A decrease of the dust temperature
with distance from the center is detected in most cores.
The obtained dust temperature profiles in most cases are close to linear ones.
Using a simple spherically-symmetric model of dust cloud,
it is shown that temperature profiles similar to the observed ones
can be obtained under the assumption of the
presence of an internal source by varying
the density profile parameters and specifying $\beta$,
a power-law index of the dust emissivity dependence on frequency, as a constant.
It is shown that the dust temperature estimates
strongly depend on the chosen value of $\beta$.
It is considered how possible variations of $\beta$ in the cloud
can affect the results obtained.


\section{Introduction}

Many details of the birth and initial evolutionary phase of massive stars
($\ga 8~$M$_{\odot}$) and star clusters remain unclear to date \cite{Tan14,Motte18}.
Research is complicated by the relative rarity of such objects, large distances to them,
the presence of a hidden phase before the main sequence and rapid evolution
compared to their low-mass counterparts.
The regions of massive star and star cluster formation are associated with massive dense cores
embedded in gas-dust clouds.
In order to determine physical conditions leading to the star formation
these objects should be investigated.

One of the effective tools for studying physical parameters and structure of dense cores
are observations of optically thin thermal radiation of dust in the range from far infrared
to submillimeter and millimeter wavelenghts, on the basis of which it is possible to calculate
dust column density, estimate density structure and mass of the object,
and determine whether the core is in equilibrium or is contracting under the influence
of its own gravity.
These estimates are usually made in the approximation of a constant dust temperature on the line of sight.
The dust temperature is estimated by fitting into the data of observations at various frequencies
the modified blackbody function: $I\propto \nu^{\beta}\,B_{\nu}(T_{\rm DUST})$,
where $\beta$ is a power-law index of the dust emissivity dependence on frequency, and
$B_{\nu}(T_{\rm DUST})$ is the Planck function at frequency $\nu$ for the dust temperature $T_{\rm DUST}$.
Dust temperature estimates depend on the $\beta$ value.
It is indicated in \cite{DL84} that if the dust is consisting of graphite and silicate particles,
the prefered value of $\beta$ at far infrared and longer wavelengths is 2.
However, the $\beta$ values obtained from observations
can be either greater or less than this value (see, e.g., Introduction to \cite{Sreenilayam14}).
For example, for a large sample of cold clumps observed with the $Planck$ telescope,
including clouds, filaments, star-forming and starless cores,
$\beta$ varies in the range $\sim 1.4-2.8$ \cite{PC11}.
In dense cores associated with star-forming regions, $\beta$ more frequently takes values
in the range $\sim 1.5-2$ (e.g., \cite{Rigby18,Sadavoy16,Yi18}).
The observed variations of $\beta$ are mainly associated with the process of growth of dust particles
due to coagulation and accretion of molecules on dust grains during evolution of the core
with increasing density and decreasing temperature, as a result of which $\beta$ decreases
(see Introduction to \cite{Kohler15} and references therein).
In the vicinity of young stars, reverse processes can also take place.
Observations also indicate the existence of an anticorrelation between dust temperature and $\beta$
(e.g., \cite{Juvela13,Juvela15,Mannfors21}) which can be both caused by the parameters dependence
(temperature and $\beta$) on each other when fitting the modified blackbody function
into observational data in the presence of noise, and by the internal properties
of dust particles themselves, as is evidenced by the results of laboratory experiments
(e.g., \cite{Boudet05,Coupeaud11}).
Variations of $\beta$ are also found in the interior of individual cores (e.g., \cite{Chen16,Bracco17,Chacon19,Galametz19}).
In protostellar cores, $\beta$ is usually reduced in regions with enhanced temperature.
At the same time, there is an evidence of a positive correlation between $\beta$ and density
(e.g., \cite{Juvela15,Tang21}) which, in particular, can indicate a slowdown in the growth of dust particles
and even the possibility of their destruction, e.g., in regions with high density
and high level of turbulence \cite{Hirashita09}.

Due to the uncertainties of the $\beta$ value, the analysis of the spatial distributions of dust temperature
is carried out less frequently than the analysis of the spatial distributions of column and volume density,
in particular, in the regions of massive star formation.
Large-scale dust temperature distributions recently obtained for a number of Galactic regions
from the $Herschel$ data in the far infrared range for $\beta=2$  \cite{Marsh17},
have a moderate angular resolution, which in many cases is insufficient for a detailed study of dust temperature
variations in the regions of massive star formation having fairly large distances.

Previously, we studied a sample of dense cores associated with the regions of massive star and star cluster formation
in various molecular lines and in continuum at a wavelength of 1.2~mm \cite{Pir03,Pir07}.
These regions contain massive dense cores and are associated with IRAS sources having high
bolometric luminosities ($\sim 10^4-10^5~L_{\odot}$), water masers \cite{Braz89},
methanol masers, and compact radio sources \cite{Walsh97,Walsh98,Pestalozzi05,QUAD}, which are indicators
of the star formation process.
These objects were also observed in various molecular lines, including dense gas indicators
(e.g., \cite{Zin95,Zin2000,Juvela96,Lap98,Harju98,Liu16}).
Analysis of observations of dust emission at 1.2~mm showed that the density profiles in these objects
are well described by either power functions with an index close to 1.6 on average,
or by profiles corresponding to the modified Bonnor-Ebert model \cite{Pir09}.

This paper presents the results of observations at a wavelength of 350~$\mu$m of eight objects
from the southern hemisphere associated with the regions of massive star and star cluster formation
from a sample observed at a wavelength of 1.2~mm \cite{Pir07}.
The observations were conducted with the APEX-12m telescope.
From the spatial distributions of intensity ratios at two wavelengths, spatial distributions
of the average dust temperature on the line of sight are obtained in objects.
The paper analyzes dust temperature profiles in the cores, makes comparison with the results of model calculations,
and discusses how possible variations of the $\beta$ index in cores can affect the results obtained.

\section{Observations}

Observations of a sample of eight star-forming regions at 350~$\mu$m
were carried out with the APEX-12m telescope in August 2010 (project 0.86F-9306A)
\footnote{This publication is based on data obtained from the telescope Atacama Pathfinder Experiment (APEX).
APEX is the result of cooperation of the Max-Planck-Institute f\"ur Radioastronomie,
the European Southern Observatory (ESO) and Onsala Space Observatory (OSO).}
with the Saboca 37-channel submillimeter bolometer (Submillimeter APEX Bolometer Camera \cite{Siringo10}).
The central frequency of the Saboca bolometers was $\sim 850$~GHz ($\sim 350$~$\mu$m), the bandwidth was $\sim 120$~GHz.
The size of the main beam of APEX-12m at this frequency is $\sim 7.5''$.
Mars, Carina, B13134, IRAS16293 and G5.89 were used for pointing.
Focusing was carried out with Mars. Errors in absolute calibration reached $\sim 30$\%.
The optical depth in the zenith was $\sim 0.9-1.2$.
The observation time was $\sim 6-12$~min, depending on the source.
Data processing was carried out using the BoA package \cite{Schuller12}.
The miniCRUSH package was also employed for comparison \cite{CRUSH,Kovacs08}.
The resulting maps were reduced to the angular resolution of $1.5''$ (the size of one pixel).
The r.m.s. deviations for residual noise on the observed maps were $\sim 5-14$~Jy/beam.
For G~351.41$+$0.64, the r.m.s. deviation was $\sim 20$~Jy/beam.

\section{Sample objects and obtained maps}
\label{objects}

The source list is given in Table~\ref{list}.
It includes the coordinates of the central positions, distances to the sources and associations with other objects.
The values of distances in some cases differ from the estimates given in \cite{Pir07} (mainly kinematic distances),
and correspond to new and/or corrected data.
In cases where the distance errors are reliably determined, these data are also indicated in Table~\ref{list}.

Fig.~\ref{maps1} shows maps of the cores at 350~$\mu$m.
For comparison, maps at 1.2~mm \cite{Pir07} are also shown.
Observations at 350~$\mu$m having significantly better angular resolution,
allowed to identify structures with smaller scales in some objects, apparently associated
with the presence of closely located compact cores or clumps that are not resolved at 1.2~mm
(e.g., in G~268.42$-$0.85, G~291.27$-$0.71 and G~294.97$-$1.73(1)).
In addition, distinct cores resolved at both wavelengths
were observed in the G~285.26--0.05, G~294.97--0.71 and G~316.77$-$0.02 regions.
A preliminary comparison revealed the presence of shifts between the 350~$\mu$m and 1.2~mm maps,
the cause of which has not been clarified yet.
From the results of fitting by two-dimensional Gaussians, exact values of the shifts were determined.
They lie within one pixel of the 1.2~mm maps (8$''$).
When constructing maps in Fig.~\ref{maps1} and in the further calculations of dust temperatures,
these shifts were taken into account.
For G~291.27$-$0.71, where the 350~$\mu$m and 1.2~mm morphologies
significantly differ from each other, no Gaussian fitting was done, and the maps were matched
approximately based on visual analysis.

In the maps shown in Fig.~\ref{maps1}, the positions of close IRAS point sources are indicated.
However, they are not always associated with the cores.
The cores themselves contain maser sources, near-infrared and radio sources, which indicate
the early stages of the massive star or star cluster formation process.
A brief summary on individual sources in the observed objects is given below.
The $SIMBAD$ \cite{Simbad}, $VisieR$ \cite{VisieR} and $maserdb.net$ \cite{Lad19} databases
are used for this compilation.

In the G~268.42$-$0.85 region, the IRAS~09002-4732 source is associated
with a young star cluster embedded in a dense gas-dust cloud.
The dominating source of the cluster is a star of the spectral class O7 \cite{Getman19}.
A water maser is observed in the region \cite{Urquhart09}.
In the 350 $\mu$m map,
two regions of enhanced intensity exist,
and the IRAS~09002-4732 source is located between them.

The IRAS~09018-4816 source in G~269.11$-$1.12 is associated with the ultra-compact H~II region \cite{QUAD}.
It is located outside the half maximum intensity regions at 350 $\mu$m and 1.2~mm
and is apparently not associated with a dense core. 
In the core, the OH \cite{Smits03} and class~II methanol masers are observed \cite{Caswell95,Urquhart15}.

The IRAS~09149-4743 source in G~270.26$+$0.83 is associated with the RCW41 H~II region,
the driving source of which is a star of the spectral class B0~V \cite{Neichel15}.
As in G~269.11$-$1.12, the IRAS source lies outside the half maximum intensity region at 350~$\mu$m and 1.2~mm,
although the position uncertainty is quite high.
In the core, a near infrared source \cite{Ortiz07}, as well as water and class~II methanol masers
\cite{Breen11,Caswell95} are observed.

The IRAS~10295-5746 source in the G~285.26$-$0.05(1) core 
is associated with the ultra-compact H~II region \cite{QUAD},
as well as OH \cite{Caswell04} and water masers \cite{Breen10}.
In the G~285.26$-$0.05(2) core, a water maser \cite{Breen10} is also observed, and 
the Br$\gamma$ line emission is detected (as in the core 1) \cite{Barnes13},
indicating the existence of an H~II region, probably immersed into a dense dust shell.

The G~291.27$-$0.71 region is located in the central part of the extended gas-dust filament.
The H~II region \cite{Eswaraiah17}, whose interaction with a parent cloud apparently determines
the morphology of a clump consisting of several fragments and corresponding IR sources, is located here.

The G~294.97$-$1.73(1) core is associated with the IRAS~11368-6312 source, the ultra-compact H~II region \cite{Walsh98},
as well as water \cite{Breen10} and class~I and class~II methanol masers \cite{Voron14,Breen19,Green12}.
The core~2 in this region is also associated with a water maser \cite{Breen10} and a class~II methanol maser \cite{Caswell09}.

The G~316.77$-$0.02 region consists of several fragments immersed into extended ridge-shaped filament.
The eastern part of the filament, near which the IRAS~14416-5937 source is located, 
is associated with maser sources, the H~II region, and the ultra-compact H~II region \cite{Watkins19}.
The core~5, located in the western part of the filament and observed at 350~$\mu$m (see Fig.~\ref{maps1}$g$)
is associated with water \cite{Breen10}, OH \cite{Caswell01}, and class~I methanol \cite{Voron14} masers.

The region G~351.41$+$0.64 (NGC6334~I), 
which is a part of a giant gas-dust complex, is a region of active star formation.
It is associated with water, OH, class~I and class~II methanol masers
(e.g., \cite{Breen10,Caswell98,Voron14,Caswell04}), the ultra-compact H~II region and a cluster of young stellar objects
(see \cite{Sadaghiani20} and references therein).
The dense core contains a large number of molecular lines, indicating high gas temperatures
which is typical for hot cores (e.g., \cite{Kalinina10,Zernickel12}).

It should be noted that according to the N$_2$H$^+$(1--0) and CS(5--4) observations \cite{Pir03,Pir07},
molecular line widths in the cores far exceed the thermal ones and in most cases increase towards the center
(this dependence is especially evident for G~351.41$+$0.64).
This may be due to an increased level of dynamic activity (turbulence, systematic motions, outflows)
in the inner regions of the cores where star formation takes place.

\begin{table}[htb]
\centering
\caption[]{Source list}
\small
\vskip 1mm
\begin{tabular}{lrrcl}
\noalign{\hrule}\noalign{\smallskip}
Source    & RA (2000)                    & Dec (2000)     & $D$   & Associations \\
          & ${\rm (^h)\  (^m)\  (^s)\ }$ &($^o$ $'$ $''$) & (kpc) & with other objects \\
\noalign{\smallskip}\hline\noalign{\smallskip}
G~268.42$-$0.85  &09 01 54.3   &$-$47 43 59  & 1.7 \cite{Getman19}   & IRAS 09002-4732 \\
G~269.11$-$1.12  &09 03 32.8   &$-$48 28 39  & 2.6 \cite{Zin95}      & IRAS 09018-4816 \\
G~270.26$+$0.83  &09 16 43.3   &$-$47 56 36  & 1.3(0.2) \cite{Neichel15}  & IRAS 09149-4743, RCW 41 \\
G~285.26$-$0.05  &10 31 30.0   &$-$58 02 07  & 4.7 \cite{Zin95}      & IRAS 10295-5746 \\
G~291.27$-$0.71  &11 11 49.9   &$-$61 18 14  & 2.4 \cite{Persi94}    & IRAS 11097-6102, NGC~3576, RCW 57 \\
G~294.97$-$1.73  &11 39 12.6   &$-$63 28 47  & 1.2 \cite{Zin95}      & IRAS 11368-6312 \\
G~316.77$-$0.02  &14 44 58.9   &$-$59 48 29  & 2.7(0.5) \cite{Watkins19}  & IRAS 14416-5937 \\
G~351.41$+$0.64  &17 20 53.4   &$-$35 47 00  & 1.3(0.3) \cite{Chibueze14} & IRAS 17175-3544, NGC~6334~I \\
\noalign{\smallskip}\hline\noalign{\smallskip}
\end{tabular}
\label{list}
\end{table}

\normalsize

\begin{figure*}
\begin{minipage}[b]{0.45\textwidth}
 \includegraphics[width=\textwidth]{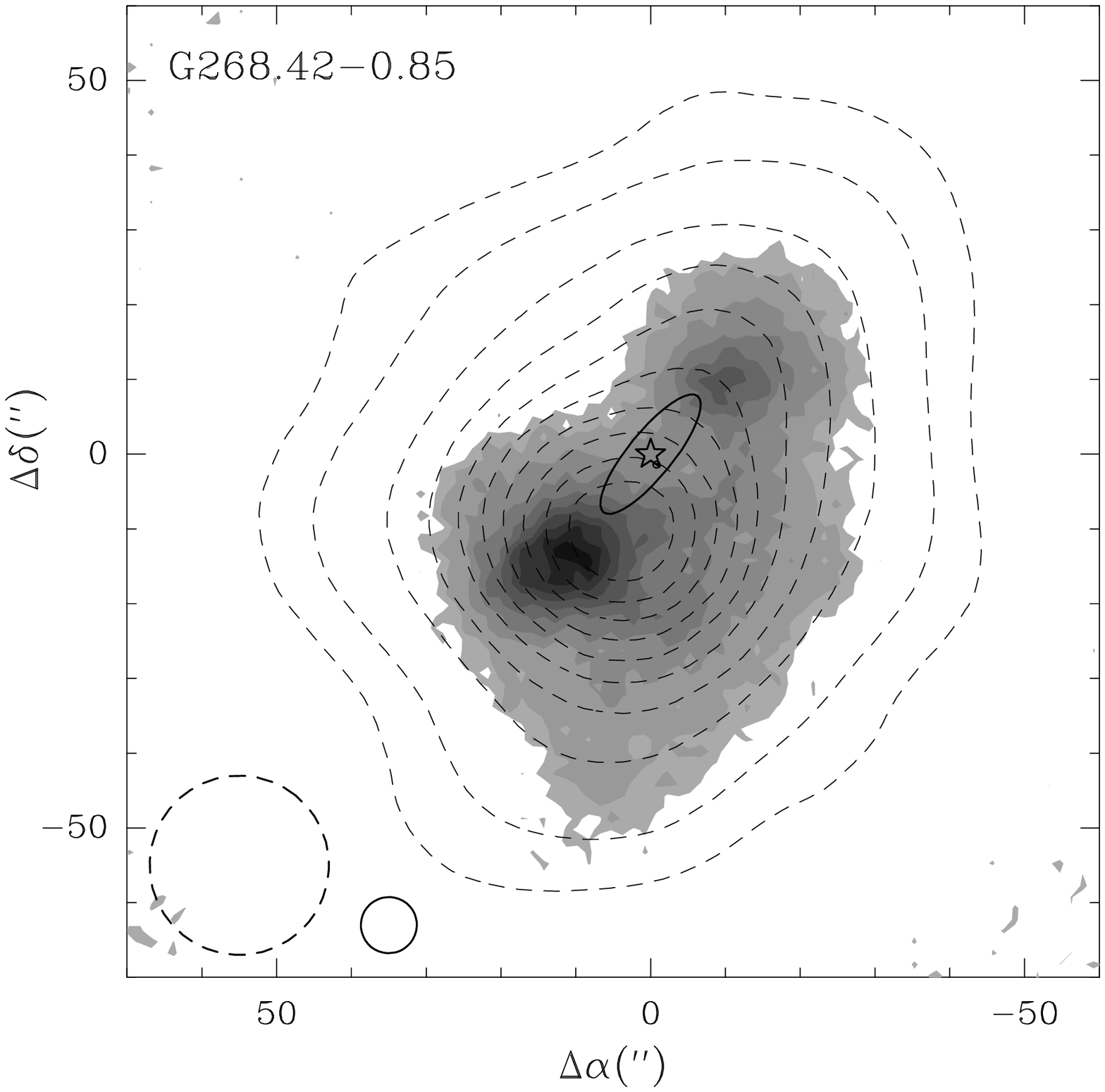}
\end{minipage}
\hspace{0.5mm}
\begin{minipage}[b]{0.45\textwidth}
 \includegraphics[width=\textwidth]{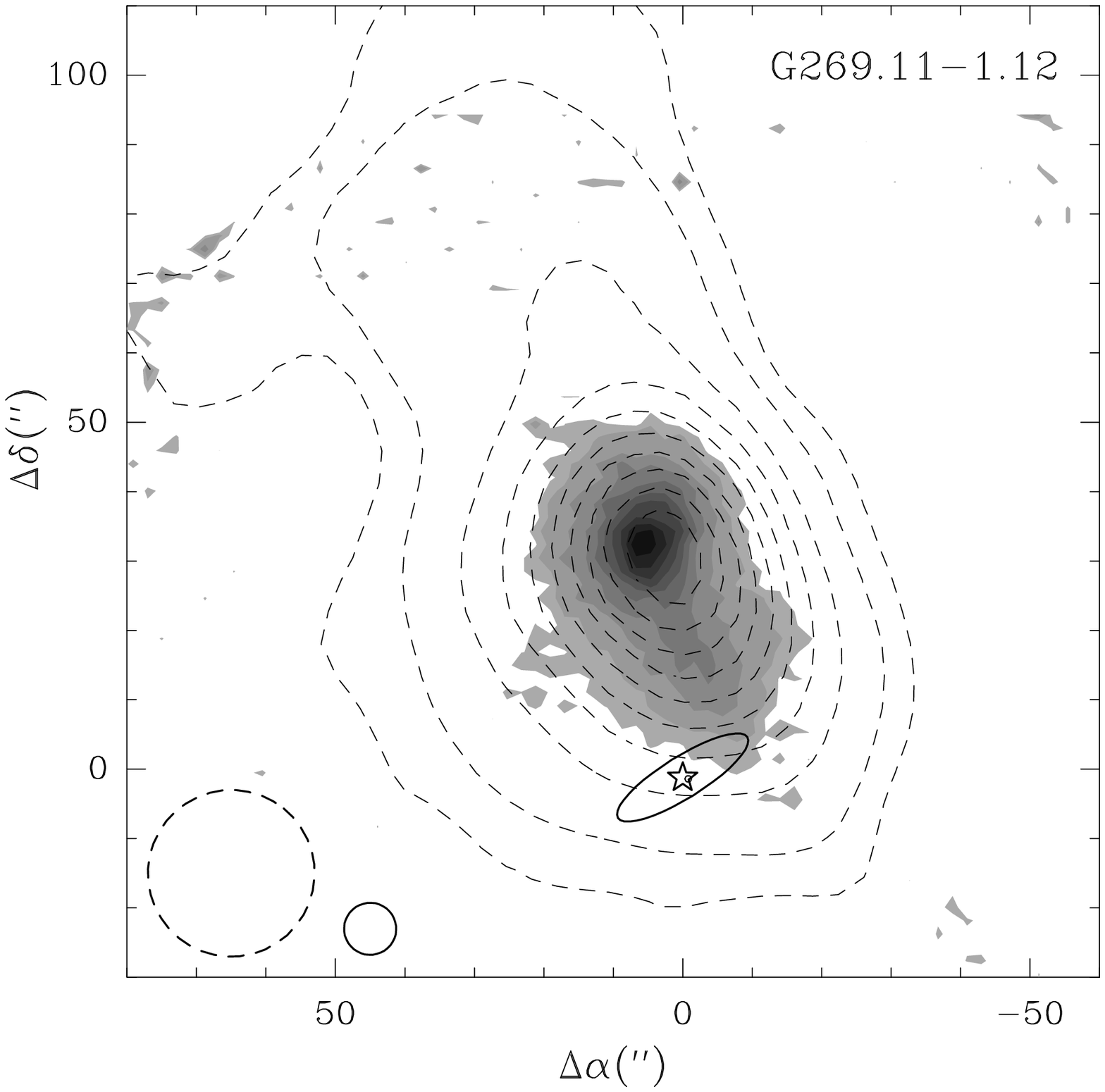}
\end{minipage}

\vskip 1mm

\begin{minipage}[b]{0.45\textwidth}
 \includegraphics[width=\textwidth]{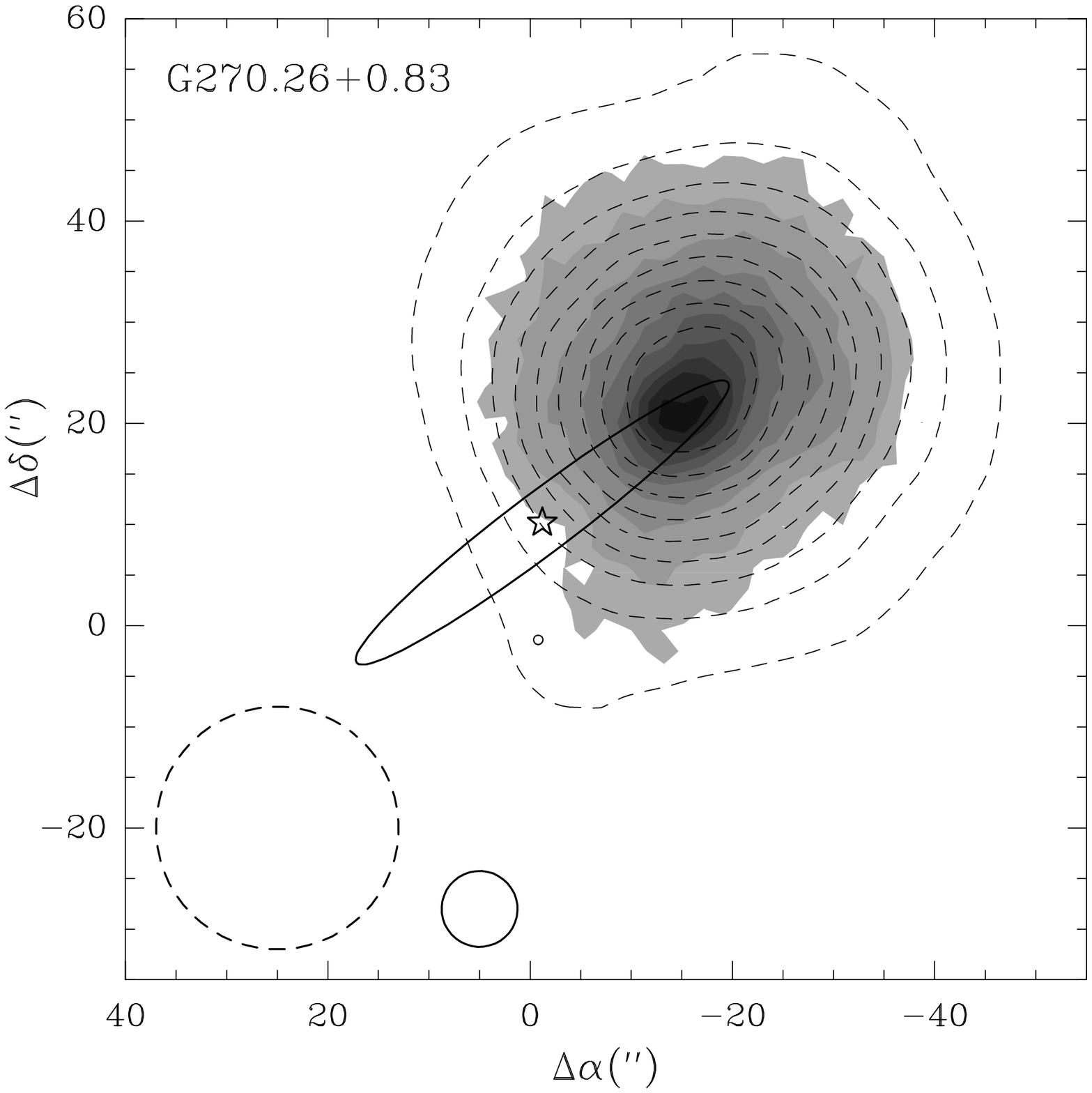}
\end{minipage}
\hspace{0.5mm}
\begin{minipage}[b]{0.45\textwidth}
 \includegraphics[width=\textwidth]{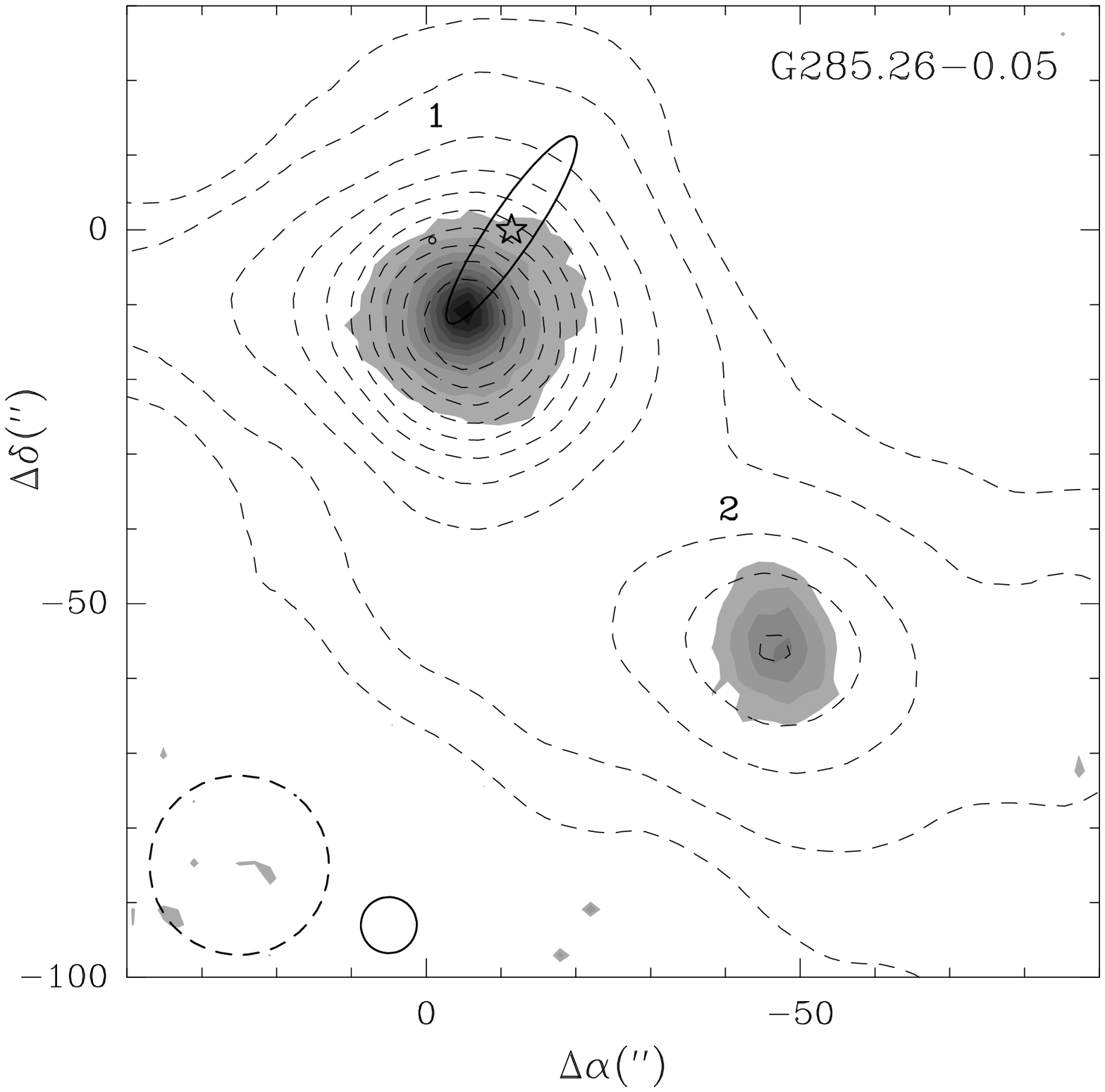}
\end{minipage}

\caption{
\footnotesize
Maps of the cores at 350~$\mu$m (grayscale) and at 1.2~mm (dashed contours).
The axes are offsets with respect to the coordinates given in Table~\ref{list}.
Grayscale and contours correspond to intensity isolines of 5\%  of the peak values and from 10\% to 90\% with an steps of 10\%.
Peak intensities at 350~$\mu$m are (in units of Jy~beam$^{-1}$): 94.3 (G268.42), 82.8 (G269.11), 93.1 (G270.26), 190.0 (G285.26),
110.0 (G291.27), 20.4 (G294.97), 22.4 (G316.77) and 430.0 (G351.41).
Peak intensities at 1.2~mm lie in the range 3.3--18.8~Jy~beam$^{-1}$ \cite{Pir07}.
For G285.26, G294.97 and G316.77, the numbers indicate the cores whose parameters were calculated separately
(the numbering corresponds to that adopted in \cite{Pir07}).
Asterisks show the positions of the IRAS point sources.
The uncertainties of these positions are shown by ellipses corresponding to the confidence level of 95\%.
In the lower left corner of each map the SEST-15m (24$''$) and APEX-12m ($7.5''$) main beams are shown.
}
\label{maps1}
\end{figure*}

\addtocounter{figure}{-1}

\begin{figure*}
\begin{minipage}[b]{0.45\textwidth}
 \includegraphics[width=\textwidth]{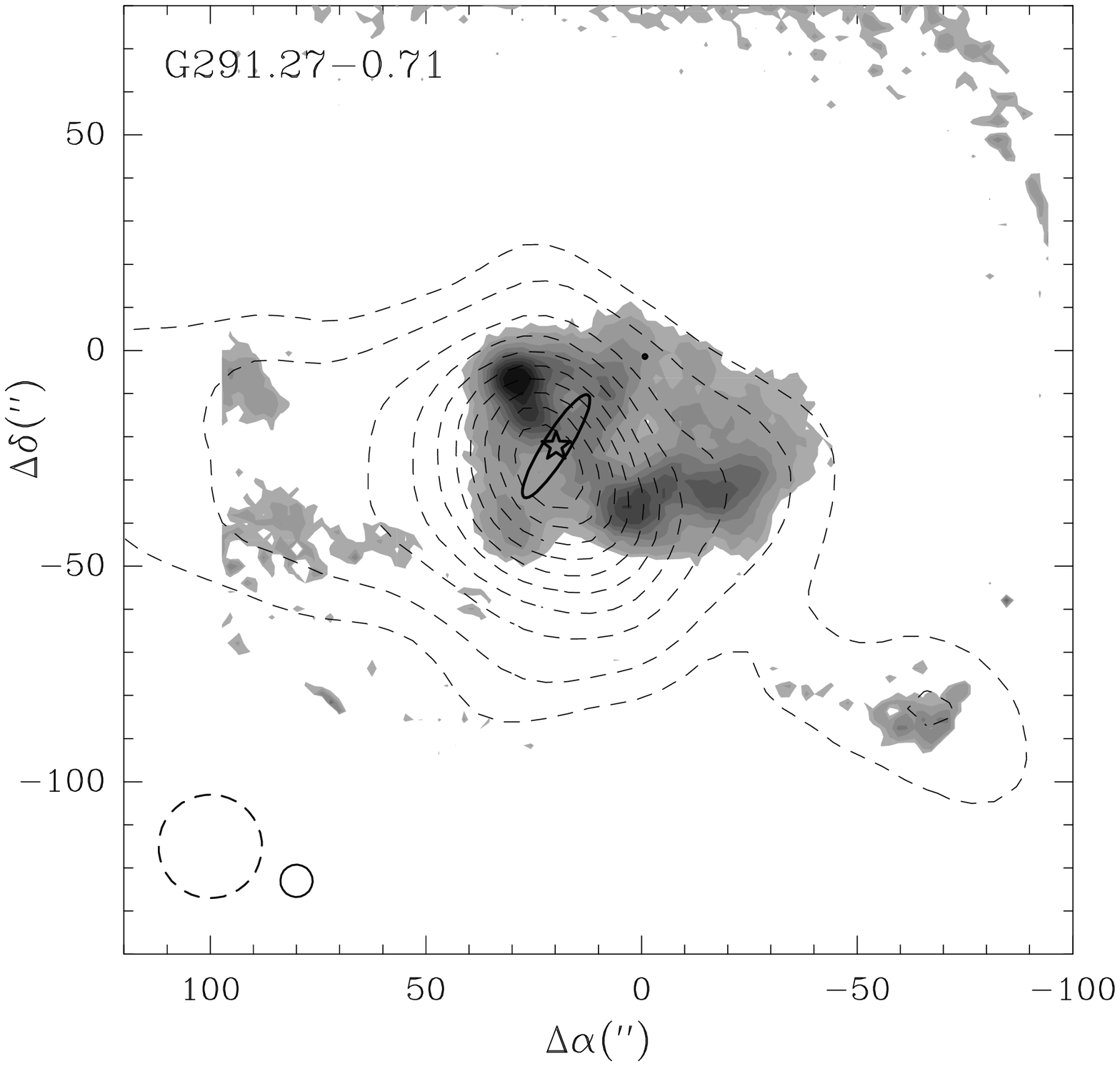}
\end{minipage}
\hspace{0.5mm}
\begin{minipage}[b]{0.45\textwidth}
 \includegraphics[width=\textwidth]{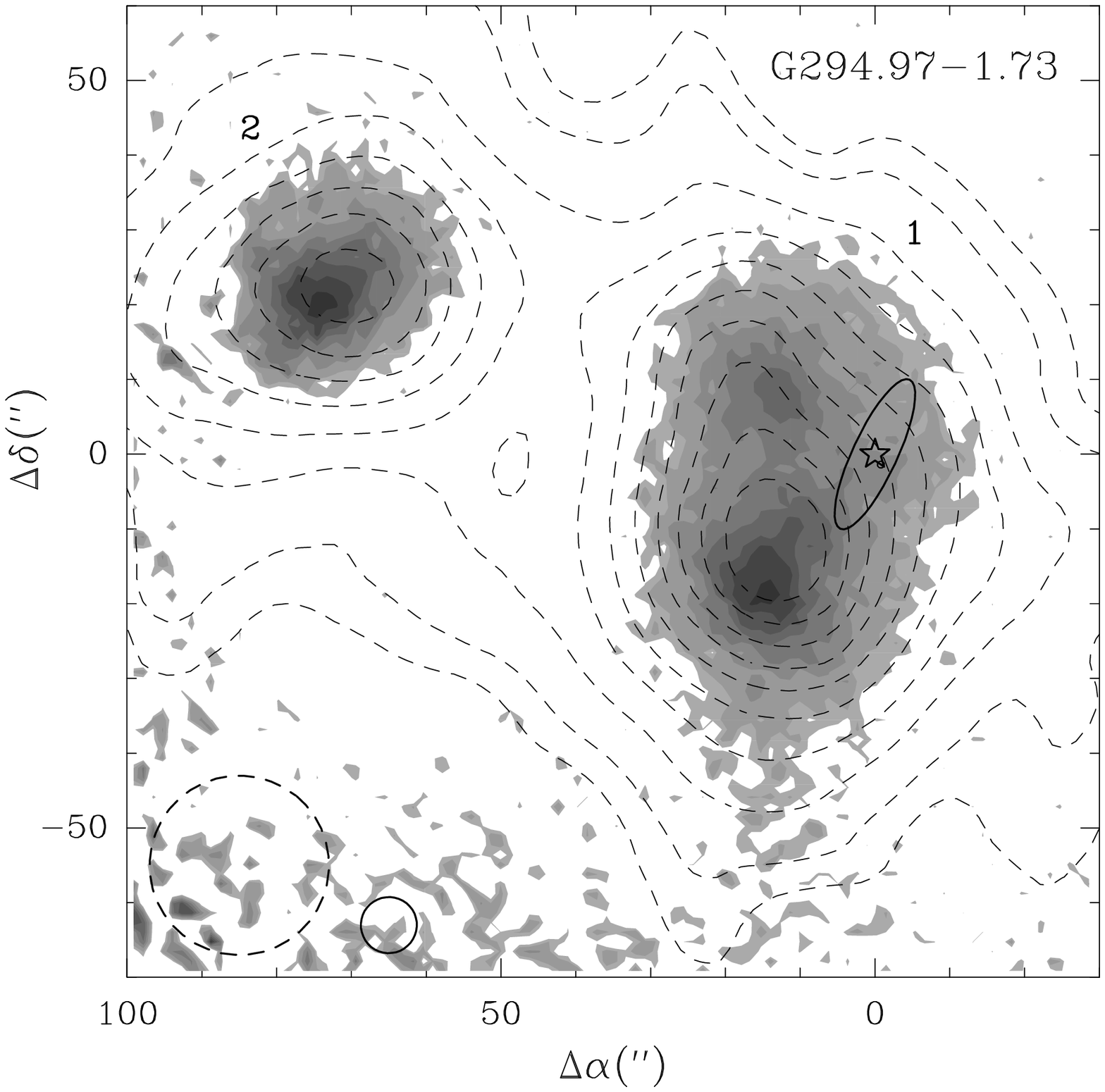}
\end{minipage}

\vskip 1mm

\begin{minipage}[b]{0.45\textwidth}
 \includegraphics[width=\textwidth]{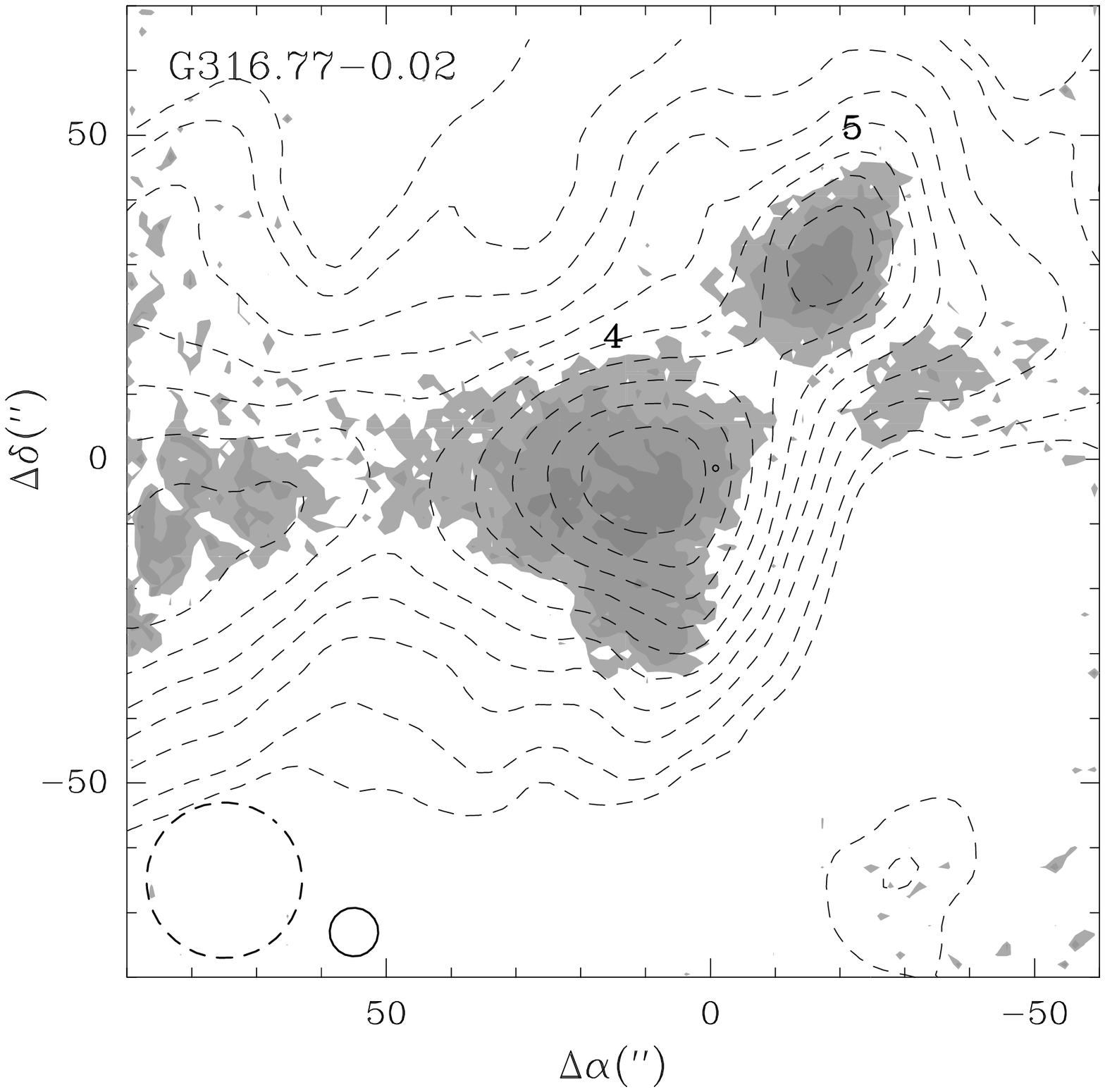}
\end{minipage}
\hspace{0.5mm}
\begin{minipage}[b]{0.45\textwidth}
 \includegraphics[width=\textwidth]{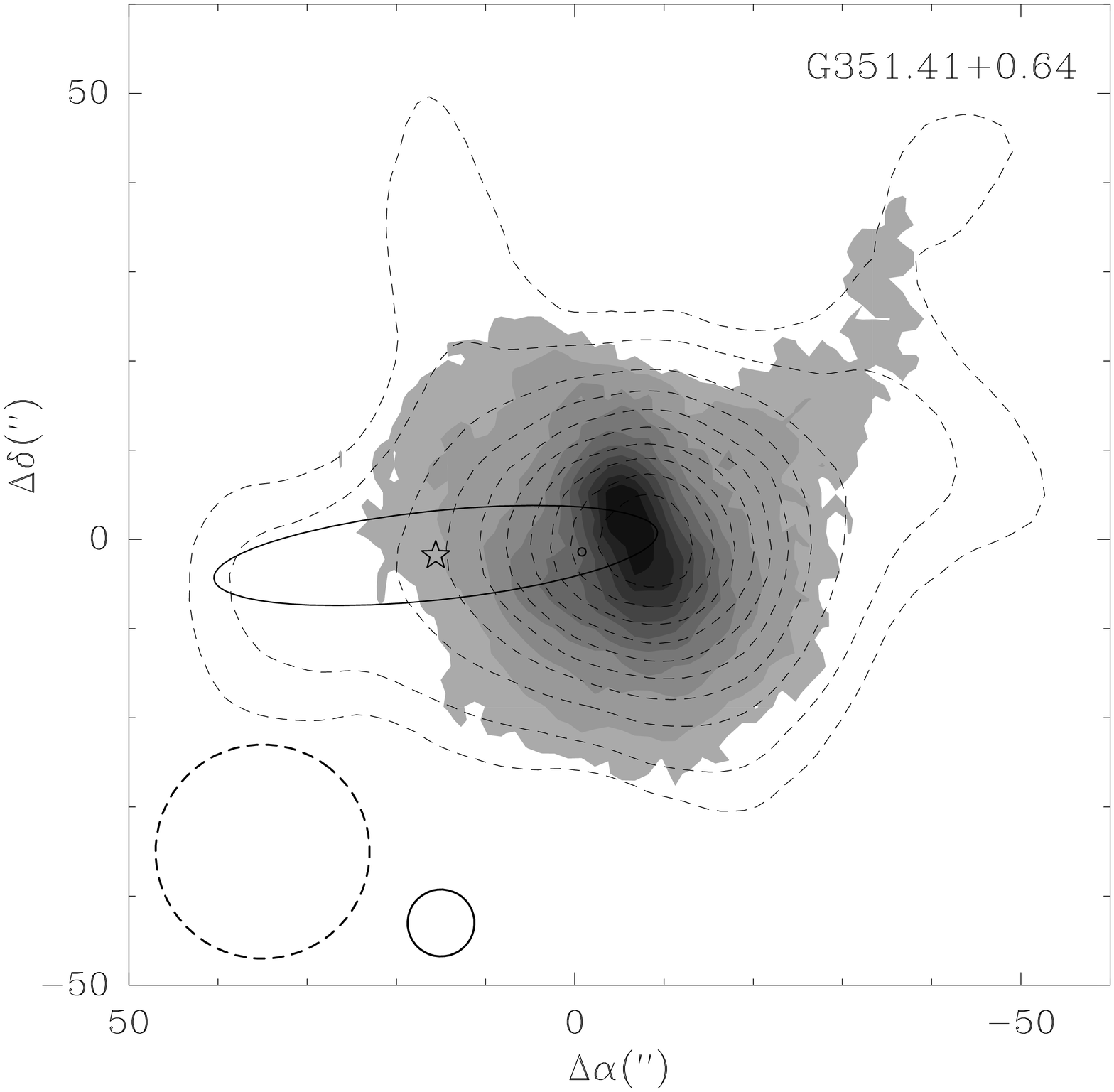}
\end{minipage}

\caption{\footnotesize Continued.}
\end{figure*}

\section{Physical parameters of the cores}

To calculate the core sizes, two-dimensional convolutions
of elliptical Gaussian with unknown parameters and a circular Gaussian with
a width equal to the telescope main beam size were fitted into the maps
or parts of the maps at 350~$\mu$m
(see \cite{Pir03} for the procedure description).
In some cases, two convolutions with different parameters were fitted simultaneously.
Angular dimensions, calculated as geometric means of the ellipse axes
at the half maximum level, and the fitting errors are given in the second column of Table~\ref{param}.
The third column gives linear core sizes ($d$), calculated
taking into account distances to the objects (Table~\ref{list}).
Their errors are mainly determined by distance errors.
For the objects whose distance errors are not found in literature,
they can apparently be quite large ($\sim 50$\% or more).
For most cores, the sizes of the emission regions at 350~$\mu$m are smaller
than the corresponding sizes at 1.2~mm \cite{Pir07}.
Besides to the effects of a narrower beam at 350~$\mu$m, this may indicate
temperature gradients in the cores (see Sec.~\ref{tprofiles}).
The fourth column of Table~\ref{param} gives the values of the total fluxes at 350~$\mu$m ($F_{350}$)
obtained by integration either over the entire observed region or over the regions occupied by individual cores
(G285.26, G294.97 and G316.77).
The fifth column gives the values of the core masses calculated in the approximation of constant
dust temperature \cite{DL94}: $M=F_{350}\,D^2/(k_{350}\,B_{350}(T_{\rm DUST}))\,R_m$,
where $k_{350}=k_{1200}$\,(1200/350)$^{\beta}$ is the dust absorption coefficient at 350~$\mu$m,
$B_{350}(T_{\rm DUST})$ is the value of the Planck function at 350~$\mu$m for the dust temperature $T_{\rm DUST}$,
which is taken to be either 25~K or 30~K based on the estimates from \cite{Pir07}, and $R_m$ is
the gas-to-dust mass ratio, taken equal to 100.
The absorption coefficient per mass unit at 1.2~mm, $k_{1200}$, is taken to be 1~cm$^2$/g
\cite{OssenkopfHenning94}, $\beta$ is taken equal to 2.
The last column gives gas mean density of the cores, calculated as the ratio of mass
to volume of a spherical core with diameter of $d$, normalized by the average
molecular mass.
For the G268.42 and G291.27 regions, densities were not calculated.

Emission region sizes at the half maximum level at 350~$\mu$m lie in the range $\sim 0.1-0.2$~pc
(except for the G291.27 region, see Table~\ref{param}), which, as is noted above,
is smaller than the corresponding sizes at 1.2~mm in most cases \cite{Pir07}.
The masses of the cores (except G291.27) lie in the range $\sim 20-1000$~$M_{\odot}$,
and the gas mean densities of the cores lie in the range $\sim 2.5\times 10^5-7.3\times 10^6$~cm$^{-3}$.
A comparison of the mass estimates with the values calculated from the 1.2~mm data \cite{Pir07}
for several objects shows their proximity within the coefficient of 2 without any systematics.
According to our estimates, the G351.41 and G285.26(1) cores have the highest mean density
(the cores with the highest IRAS luminosities of our sample \cite{Pir07}).
The cores in the G294.97 and G316.77 regions have the smallest mean density. 
Note, that these values may be overestimated, since the masses are calculated for the regions with the sizes
higher than the core sizes at the half maximum level.
At the same time, the calculated masses may be underestimated.
As the further analysis shows,
the dust temperature in the cores apparently decreases with a distance from the center,
which should lead to an increase of the mass estimates.

\begin{table}[htb]
\centering
\caption[]{Core parameters}
\small
\vskip 1mm
\begin{tabular}{lrrrrr}
\noalign{\hrule}\noalign{\smallskip}
Source           & $\Delta(\Theta)$  & $d$  & $F_{350}$ & $M$           & $n$ \\
                   & ($''$)           & (pc) & (Jy)       & (M$_{\odot}$) & (cm$^{-3}$) \\
\noalign{\smallskip}\hline\noalign{\smallskip}
G~268.42$-$0.85     & 28.4(0.3)$^a$ & 0.23        & 960  & 360 &  \\
G~269.11$-$1.12     & 17.2(0.2)     & 0.22        & 390  & 480 & 1.5\,10$^6$ \\
G~270.26$+$0.83     & 19.2(0.1)     & 0.12        & 470  & 100 & 1.9\,10$^6$ \\
G~285.26$-$0.05 (1) & 8.4(0.1)      & 0.19        & 355  & 1010& 5.4\,10$^6$ \\
G~285.26$-$0.05 (2) & 9.3(0.2)      & 0.21        & 120  & 490 & 1.7\,10$^6$ \\
G~291.27$-$0.71     & $\sim 80^b$   & $\sim 0.9$  & 1320 & 1400&  \\
G~294.97$-$1.73 (1) & 32.3(0.3)     & 0.19        & 200  & 50  & 2.5\,10$^5$ \\
G~294.97$-$1.73 (2) & 17.5(0.3)     & 0.10        & 70   & 20  & 6.3\,10$^5$ \\
G~316.77$-$0.02 (5) & 16.0(0.3)     & 0.21        & 110  & 150 & 5.5\,10$^5$\\
G~351.41$+$0.64     & 23.7(0.2)     & 0.15        & 3360 & 730 & 7.3\,10$^6$\\
\noalign{\smallskip}\hline\noalign{\smallskip}
\end{tabular}
\\
\begin{flushleft}
\small
{
$^a$ -- the size of the south-eastern core (see Fig.~\ref{maps1}a), \\
$^b$ -- the approximate size of the central "spot" (see Fig.~\ref{maps1}$e$).
}
\end{flushleft}
\label{param}
\end{table}

\normalsize

\section{Spatial distribution of dust temperature in the cores}
\label{tprofiles}

The dust temperature can be derived from the ratio of intensities of optically thin dust emission ($I$)
at two frequencies ($\nu_1$ and $\nu_2$):

\begin{equation}
\frac{I(\nu_1)}{I(\nu_2)}=\Bigl (\frac{\nu_1}{\nu_2} \Bigr)^{3+\beta}\hspace{2mm}
\frac{\exp(\frac{h\nu_2}{k_BT_{\rm DUST}})-1}{\exp(\frac{h\nu_1}{k_BT_{\rm DUST}})-1}
\hspace{1mm},
\label{eq1}
\end{equation}

\noindent{where $h$ and $k_B$ are the Planck and Boltzmann constants, respectively.}
Knowing spatial distribution of the intensity ratios at two frequencies and specifying
the $\beta$ value, it is possible to estimate the spatial distribution of dust temperature.
Since the observed intensity of dust emission is an integral over the line of sight,
the dust temperatures obtained from the intensity ratios are the values averaged
over the column on the line of sight.

\begin{figure*}
\centering
\begin{minipage}[b]{0.35\textwidth}
 \includegraphics[width=\textwidth]{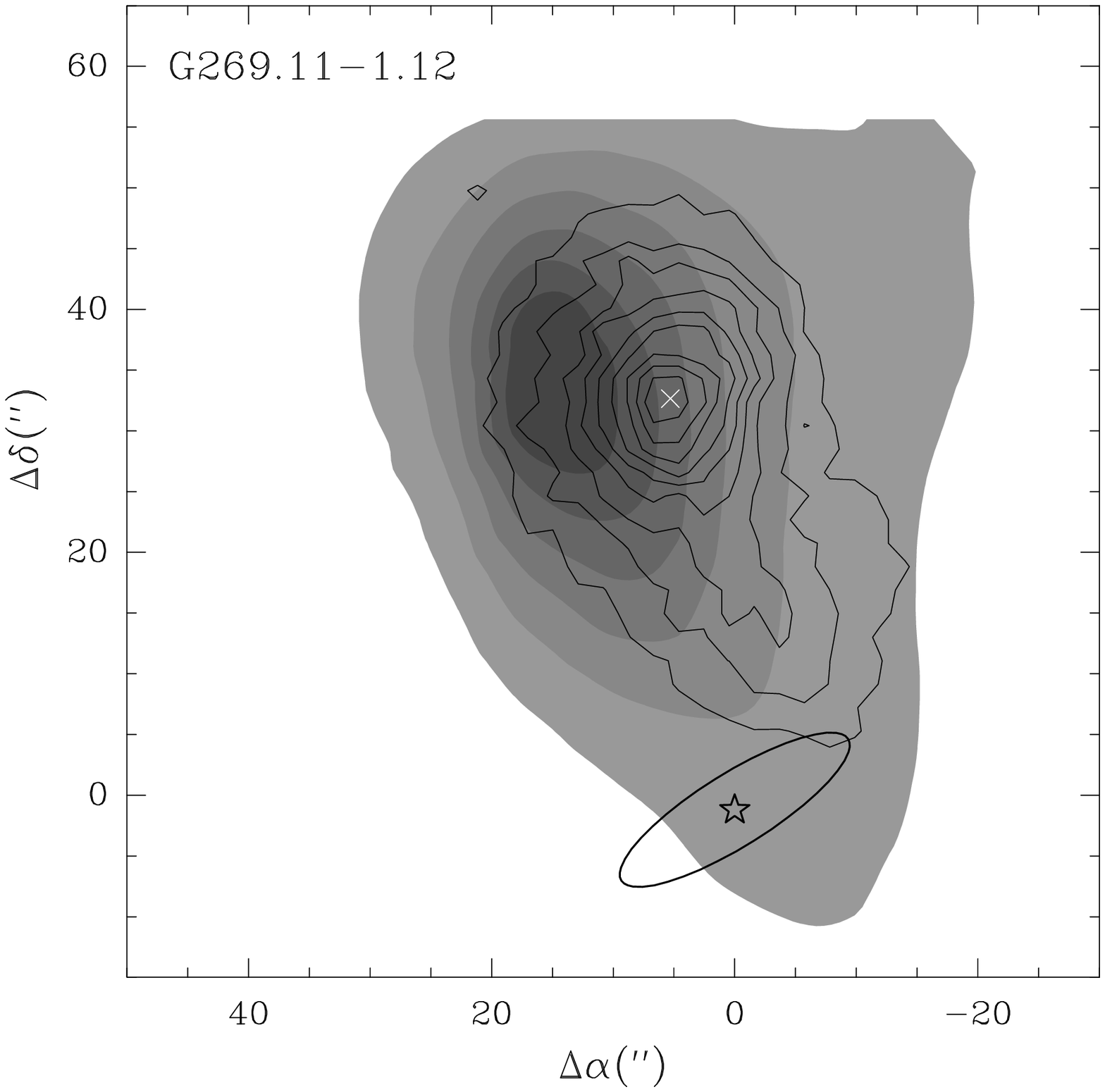}
\end{minipage}
\hspace{0.5mm}
\begin{minipage}[b]{0.35\textwidth}
 \includegraphics[width=\textwidth]{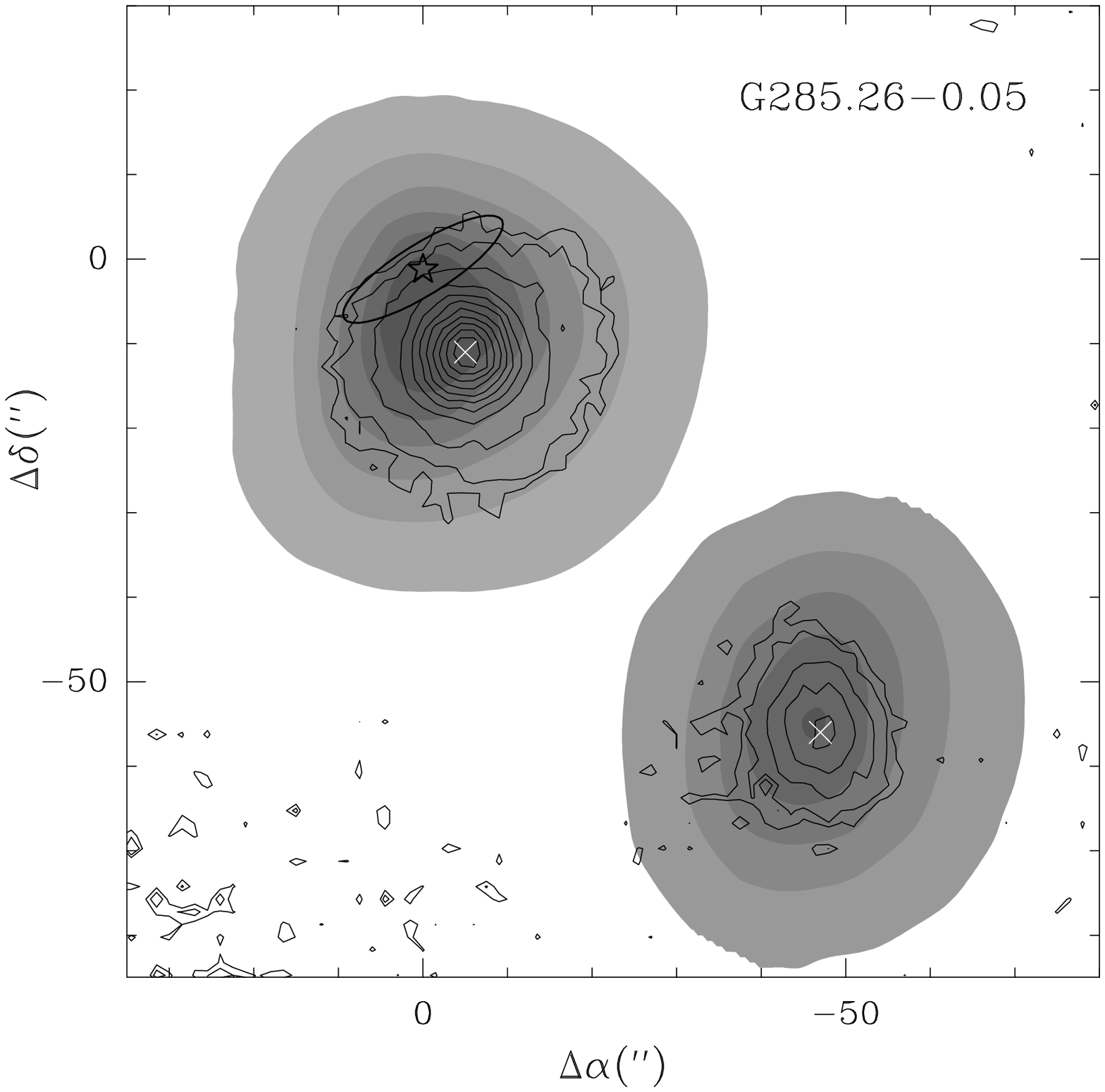}
\end{minipage}

\vskip 1mm

\begin{minipage}[b]{0.35\textwidth}
 \includegraphics[width=\textwidth]{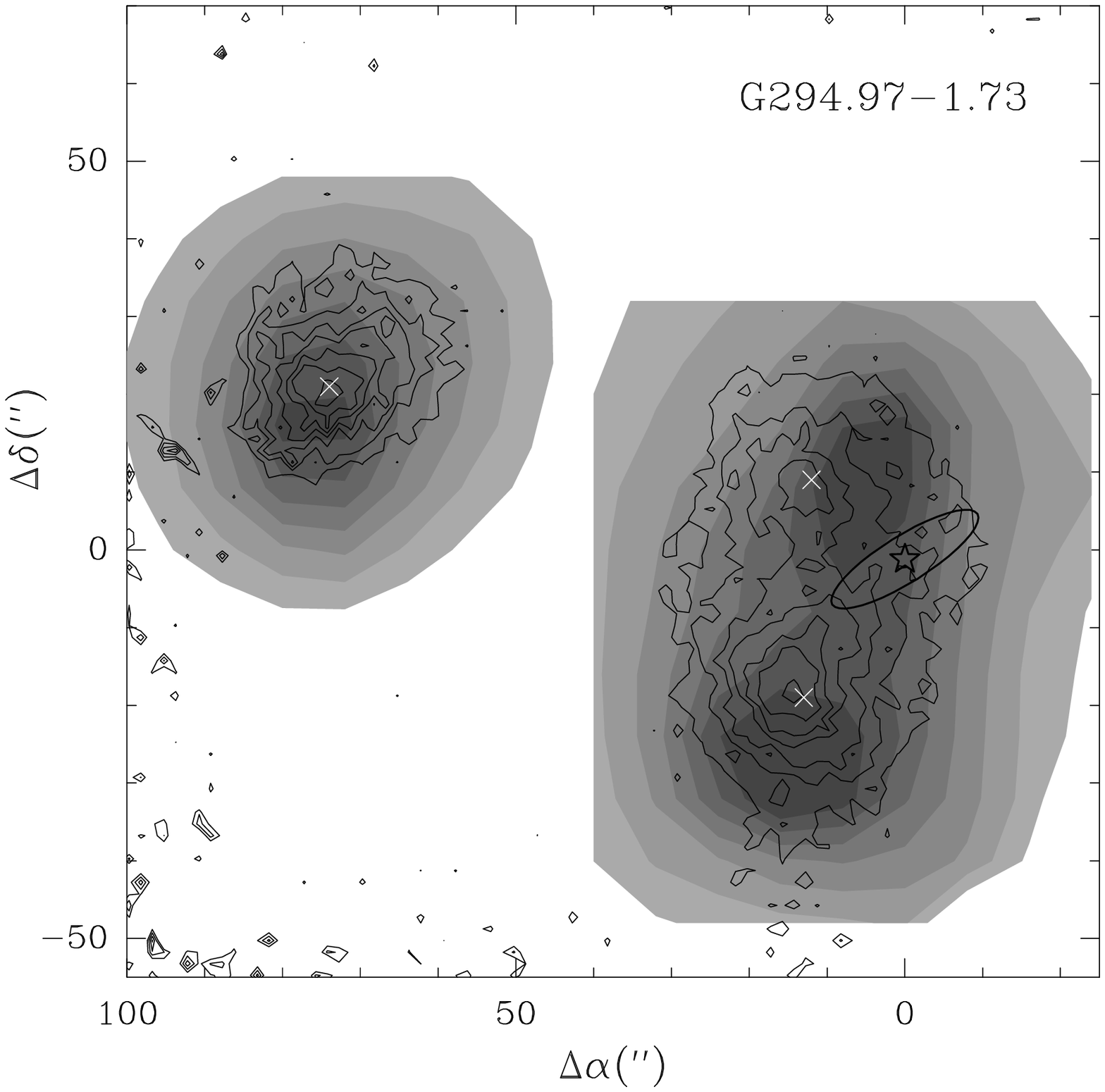}
\end{minipage}
\hspace{0.5mm}
\begin{minipage}[b]{0.35\textwidth}
 \includegraphics[width=\textwidth]{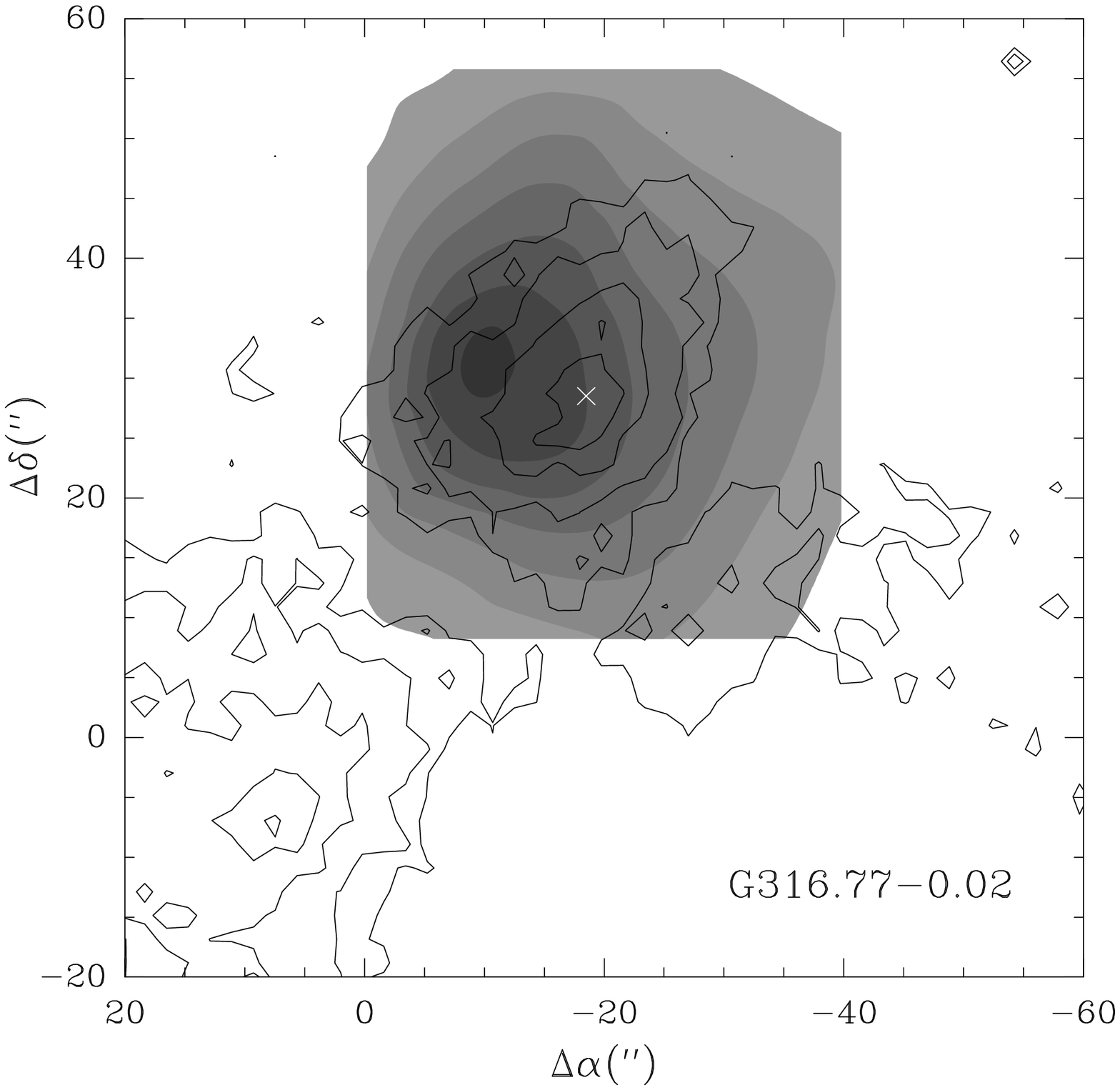}
\end{minipage}


\begin{minipage}[b]{0.32\textwidth}
 \includegraphics[width=\textwidth]{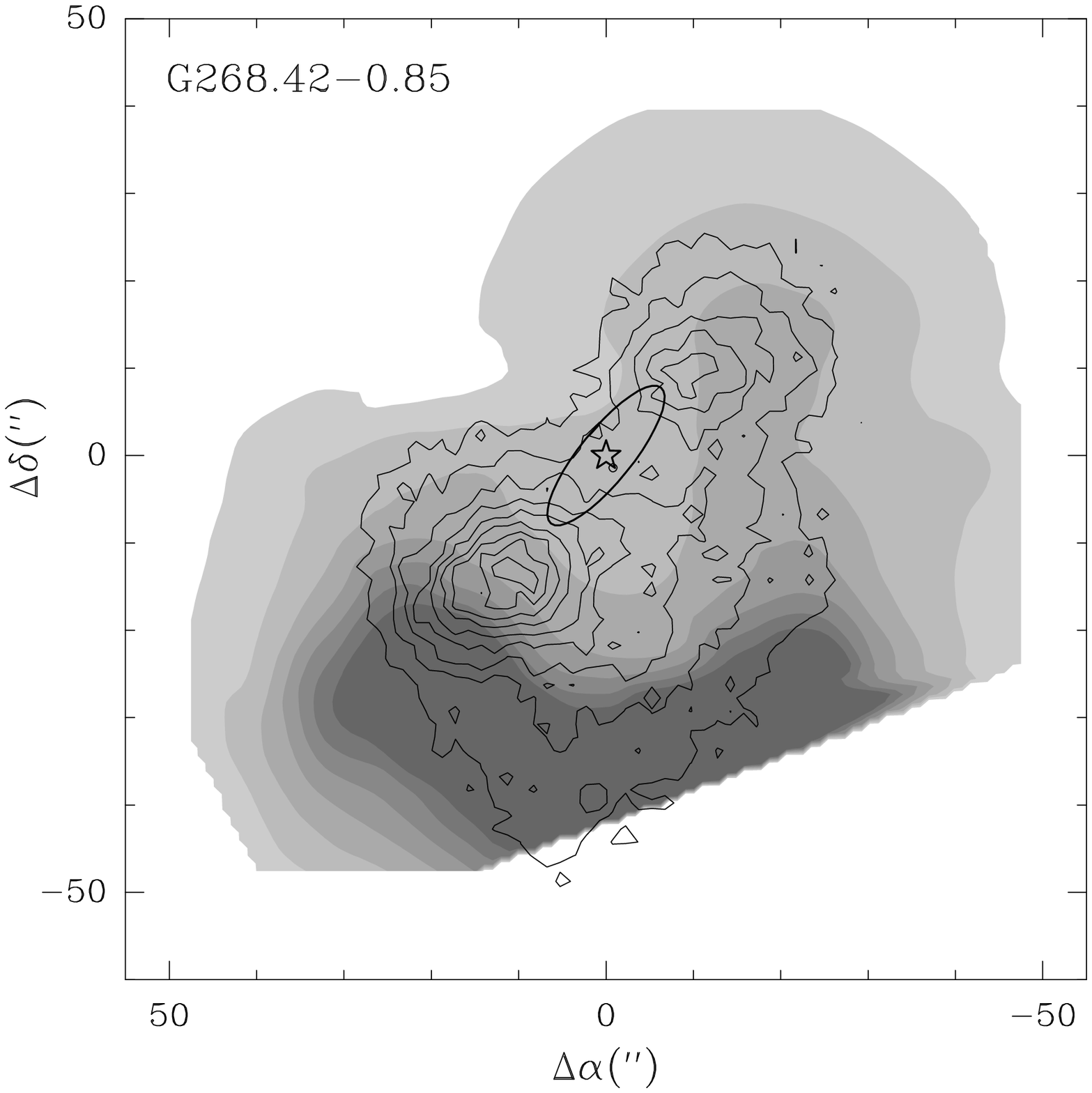}
\end{minipage}
\hspace{0.5mm}
\begin{minipage}[b]{0.32\textwidth}
 \includegraphics[width=\textwidth]{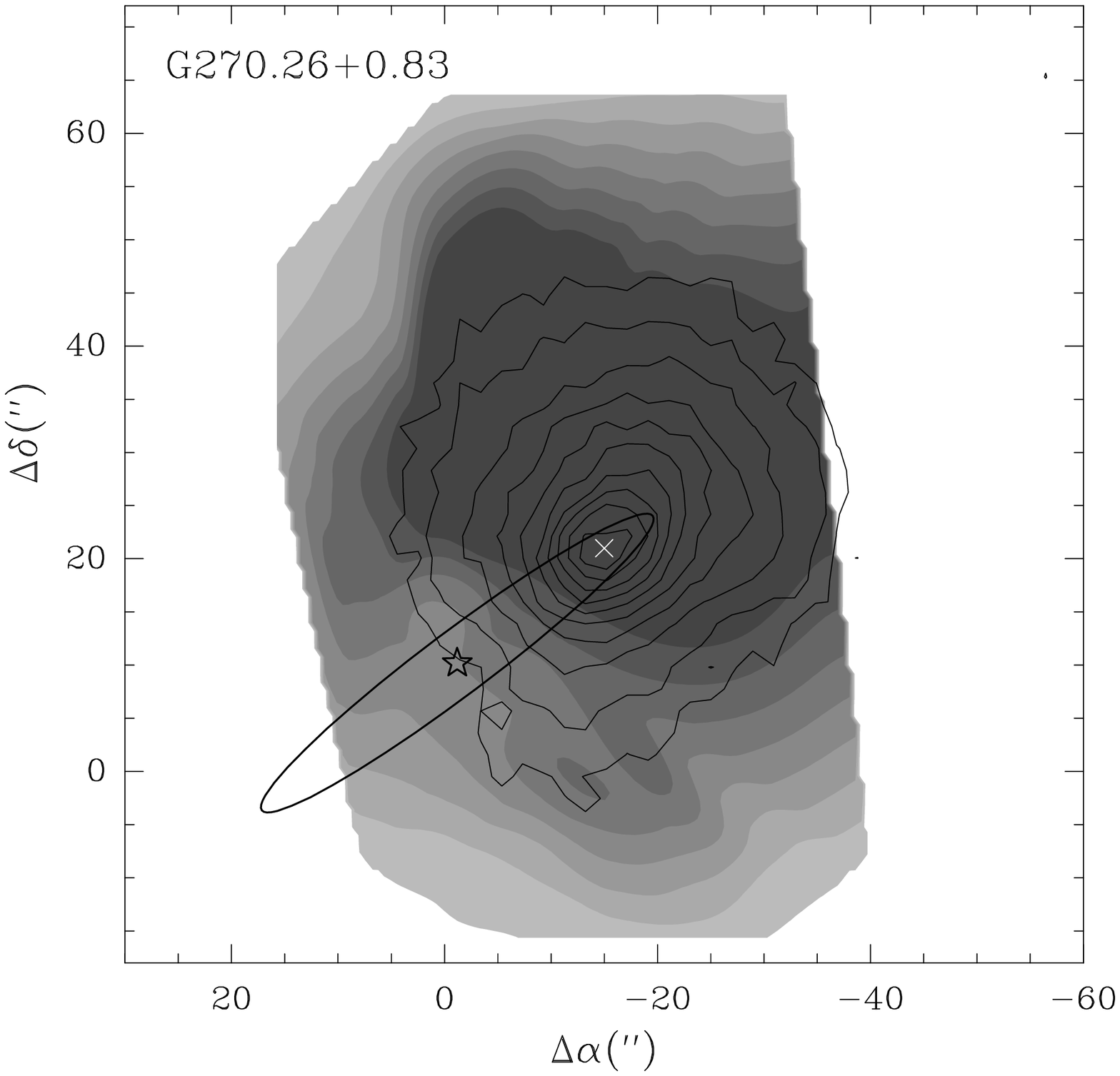}
\end{minipage}
\hspace{0.5mm}
\begin{minipage}[b]{0.32\textwidth}
 \includegraphics[width=\textwidth]{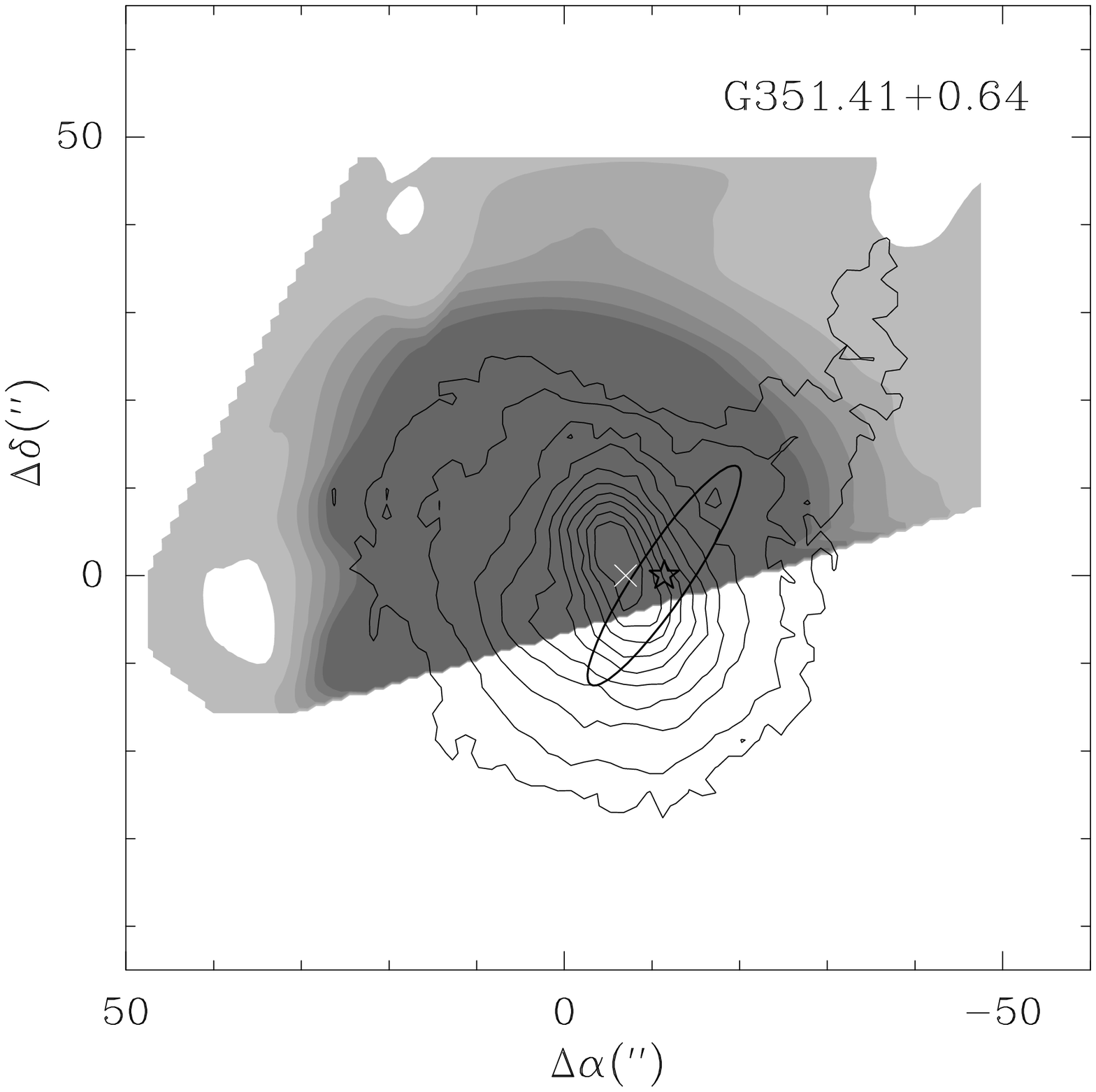}
\end{minipage}

\caption{
\footnotesize
Spatial distributions of dust temperature in the objects for $\beta=1.7$.
Grayscale corresponds to the temperature isolines from 10~K to 70~K with a step of 10~K (G268.42),
from 20~K to 70~K with a step of 10~K (G269.11),
from 10~K to 80~K with a step of 10~K (G270.26),
from 10~K to 40~K with a step of 5~K (G294.97, G316.77),
from 10~K to 60~K with a step of 10~K (G285.26, G351.41).
The contours are the 350~$\mu$m intensity maps.
The IRAS sources and their position uncertainties are indicated.
Crosses indicate the central positions used for calculations
of the dust temperature profiles.
}
\label{tdust}
\end{figure*}

First, observational data at 350~$\mu$m were convolved to the main beam size at 1.2~mm (24$''$).
The corresponding intensities and their errors were calculated for a grid that coincides 
with the 1.2~mm map grid (pixel size 8$''$) taking shifts into account.
The dust temperatures at the grid nodes are calculated from Eq.~(\ref{eq1})
for $\beta=1.7$ and 2.0.
The dust temperature errors ($\sigma_T$) were calculated taking into account the errors
of intensity ratios at two wavelengths.
For the G291.27 region, the value of the shift between the maps was not defined,
and the dust temperatures were not calculated.
For the G316.77 region, the dust temperatures were calculated for the core 5.
Figure~\ref{tdust} shows maps of the spatial distributions of dust temperature for $\beta=1.7$.
The temperatures higher than $3\,\sigma_T$ are used in the maps.

For the G285.26 and G294.97 cores, the spatial distributions of dust temperature
correlate well with the 350~$\mu$m maps,
and, hence, with total dust and gas column densities.
The centers of these maps are close to each other.
The dust temperature map in G294.97 shows that the core~1 consists of two regions, northern and southern.
For G269.11 and G316.77, the centers of the dust temperature maps are displaced with respect to the centers
of the corresponding emission regions at 350~$\mu$m.
For G268.42, the dust temperatures appear to be higher at the periphery of the observed area
and does not correlate with the 350~$\mu$m and 1.2~mm maps.
This object was excluded from the further analysis.
For G351.41, the dust temperature errors in the southern part of the core
turned out to be quite large, so the map is shown only for the northern part.

\begin{figure*}
\centering
\begin{minipage}[b]{0.3\textwidth}
 \includegraphics[width=\textwidth]{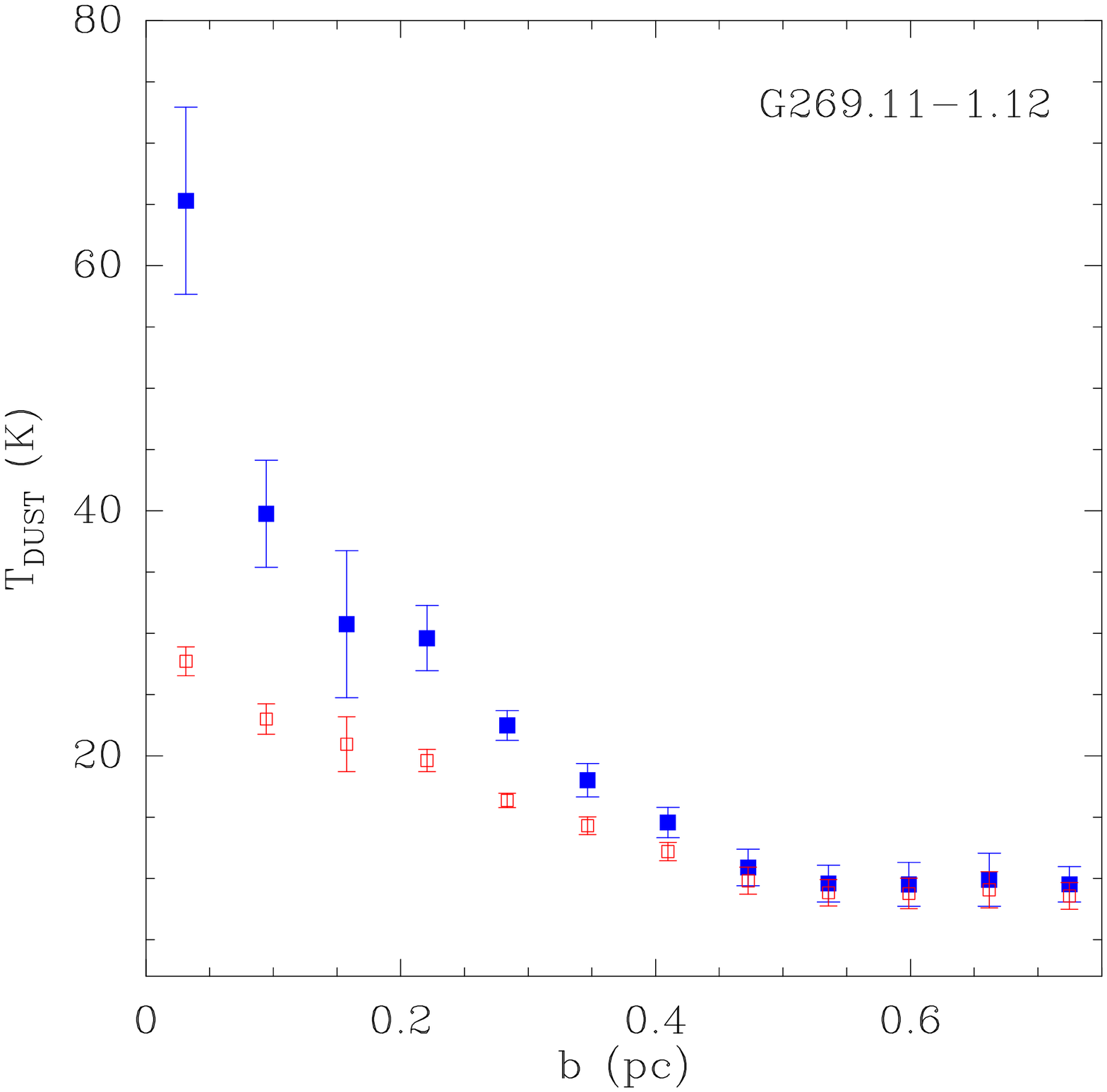}
\end{minipage}
\hspace{0.5mm}
\begin{minipage}[b]{0.3\textwidth}
 \includegraphics[width=\textwidth\centering]{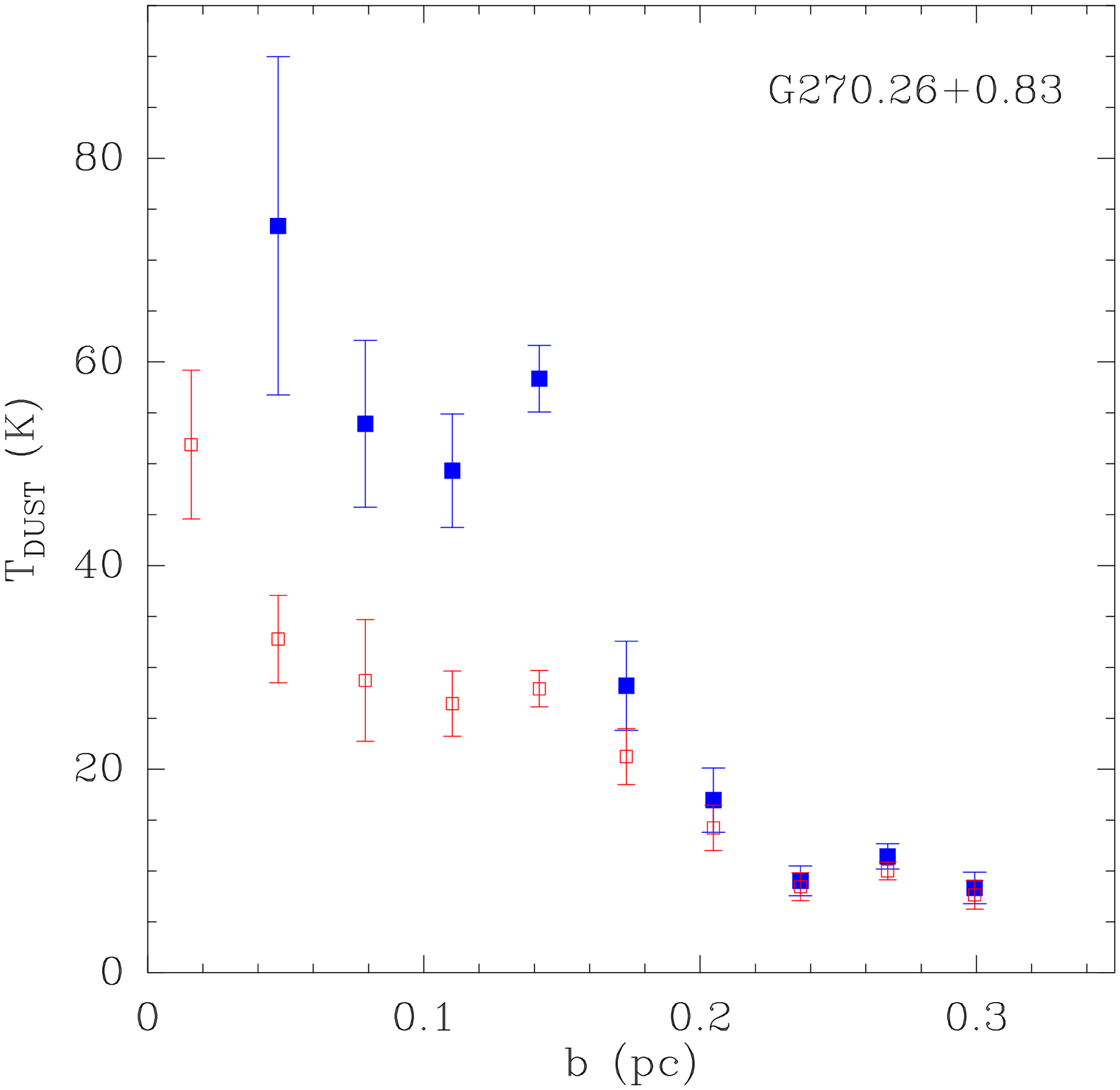}
\end{minipage}
\hspace{0.5mm}
\begin{minipage}[b]{0.3\textwidth}
 \includegraphics[width=\textwidth]{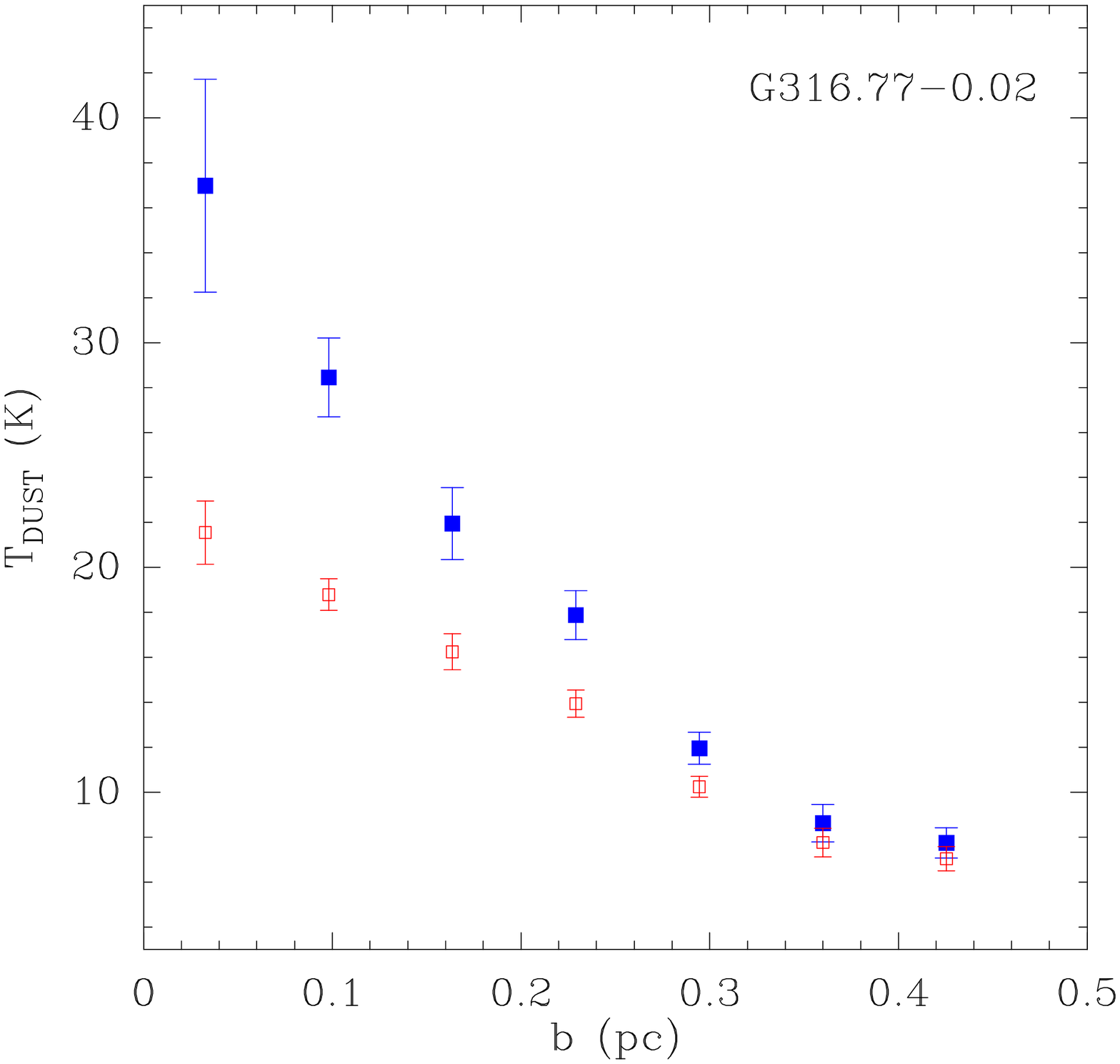}
\end{minipage}

\vskip 1mm

\begin{minipage}[b]{0.3\textwidth}
 \includegraphics[width=\textwidth]{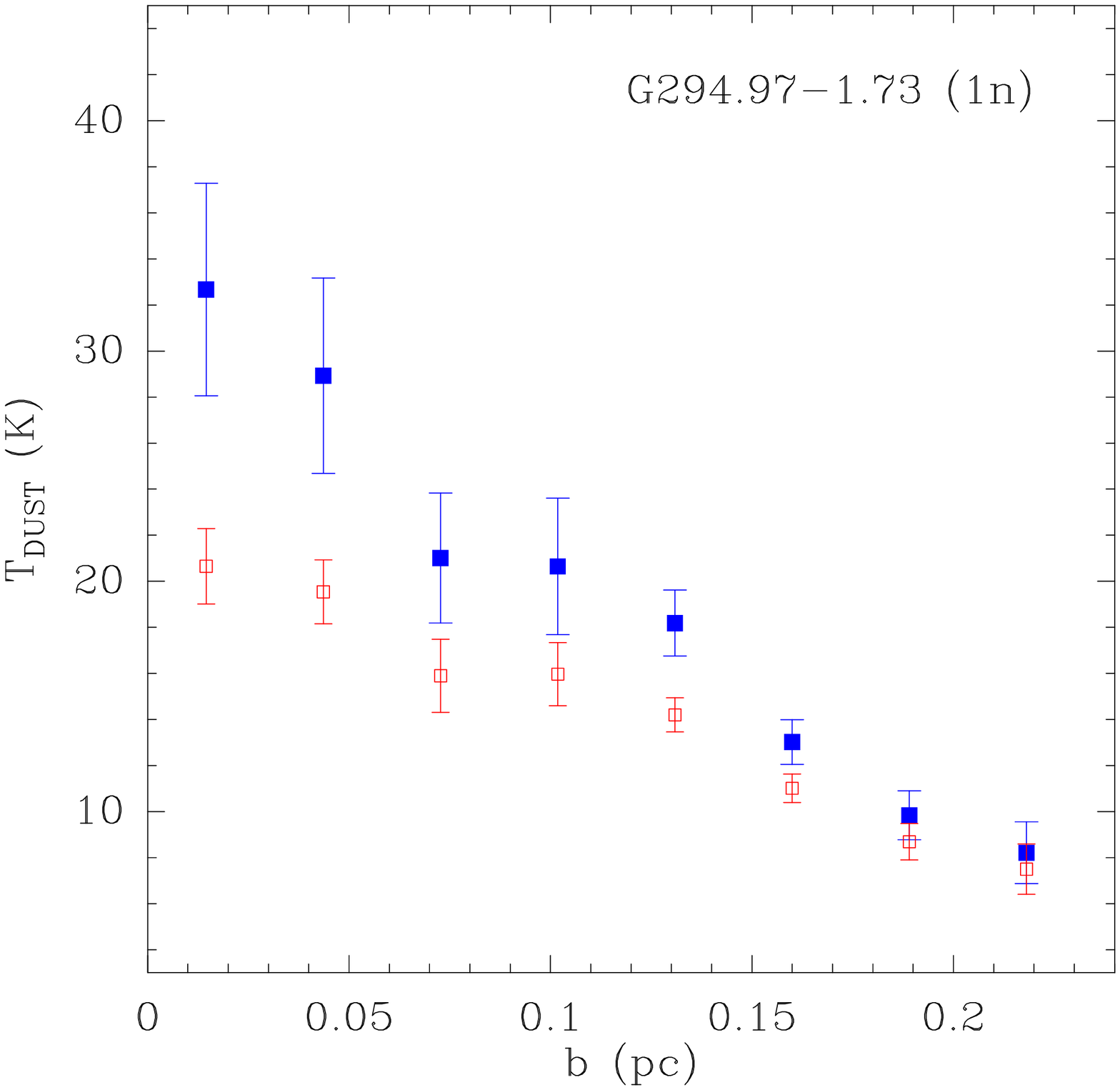}
\end{minipage}
\hspace{0.5mm}
\begin{minipage}[b]{0.3\textwidth}
 \includegraphics[width=\textwidth]{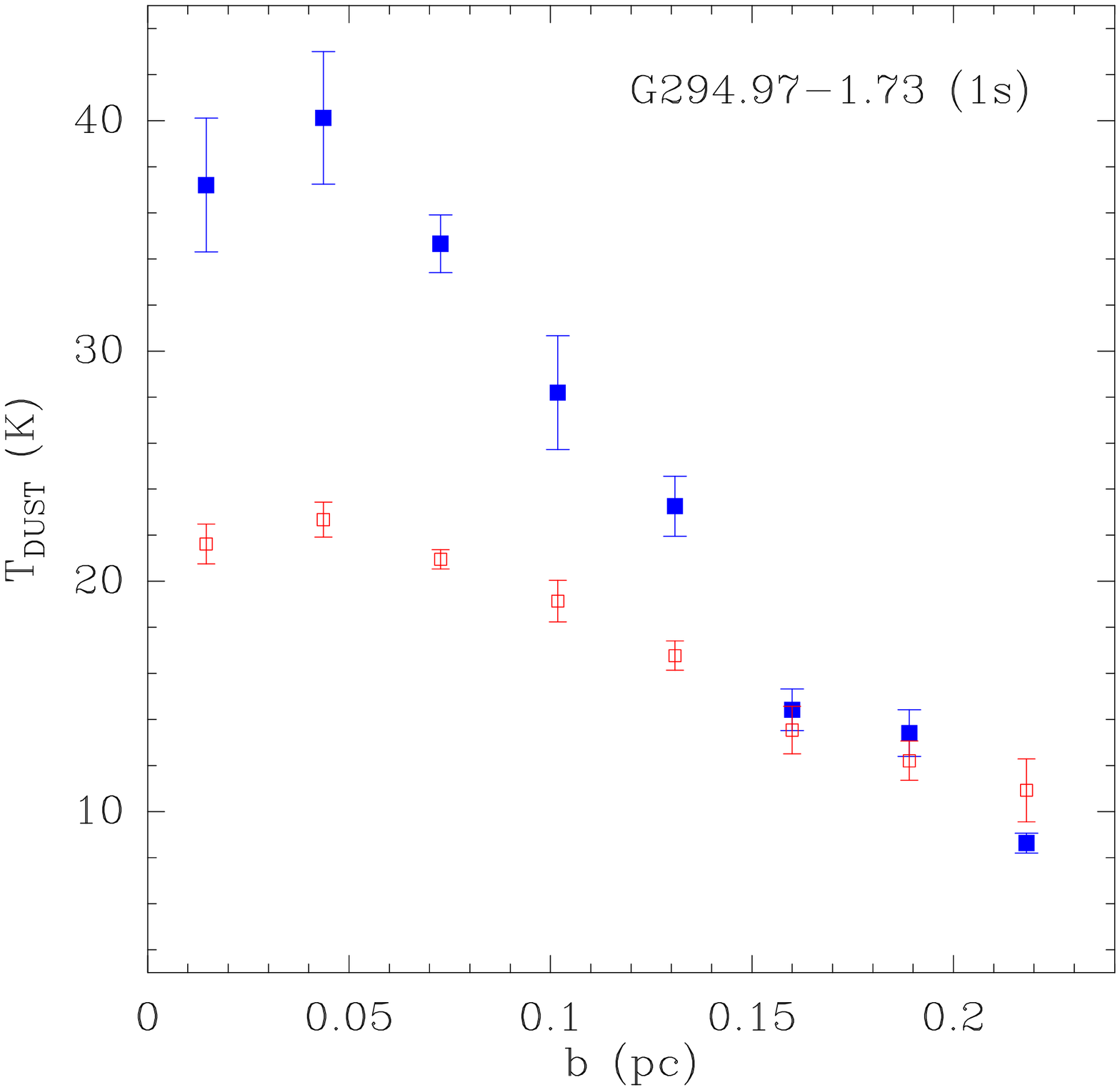}
\end{minipage}
\hspace{0.5mm}
\begin{minipage}[b]{0.3\textwidth}
 \includegraphics[width=\textwidth]{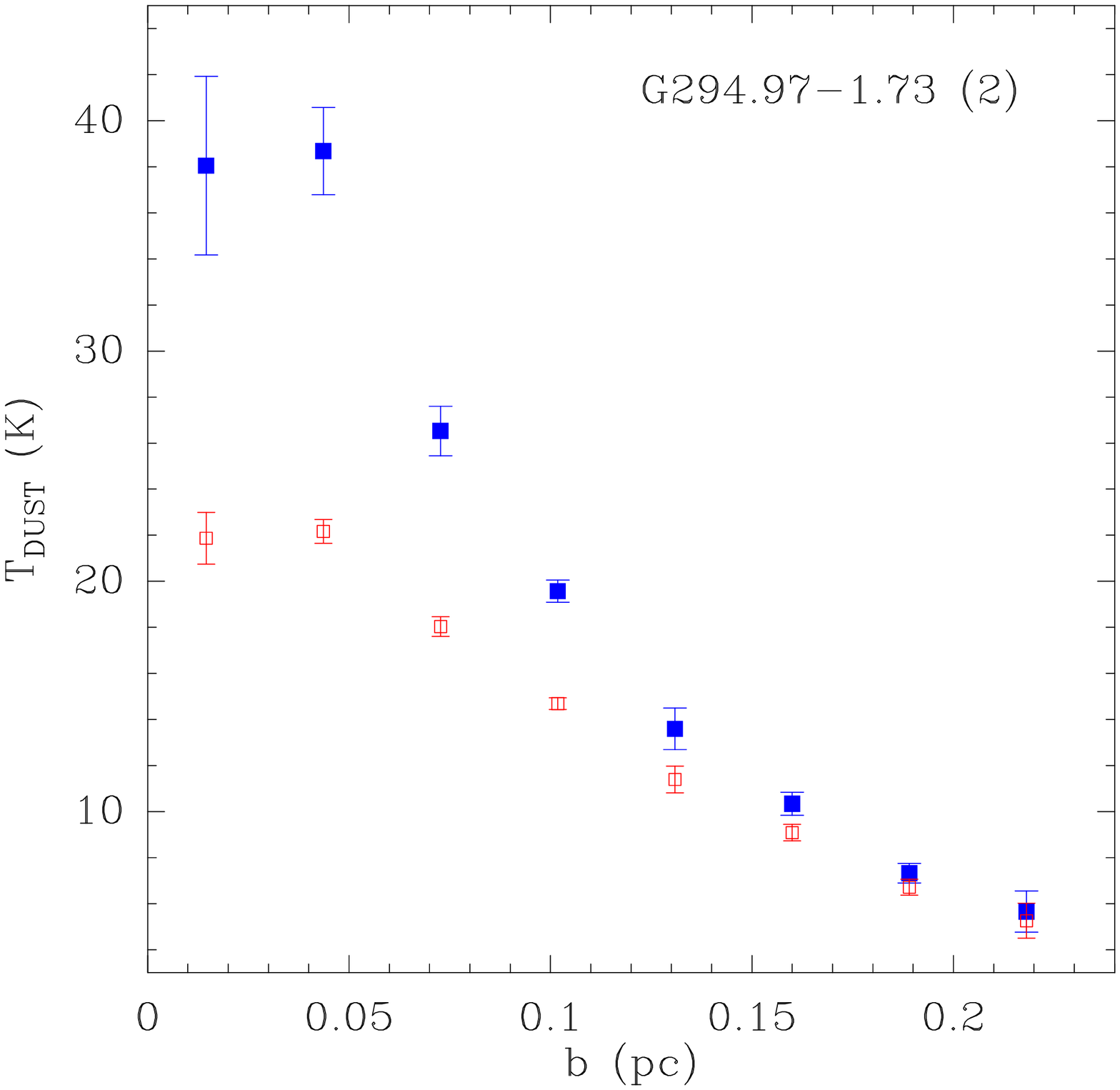}
\end{minipage}

\vskip 1mm

\begin{minipage}[b]{0.3\textwidth}
 \includegraphics[width=\textwidth\centering]{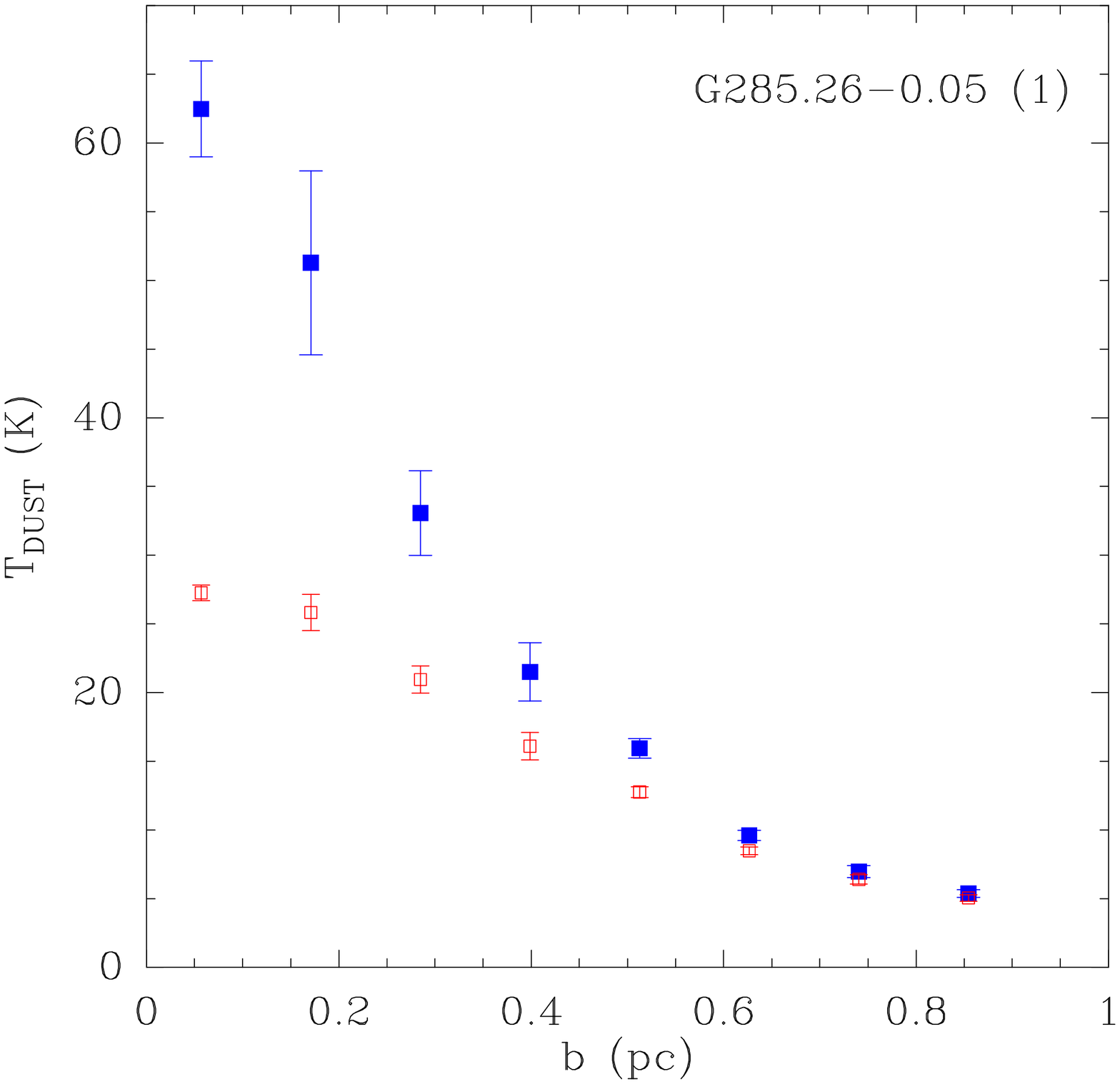}
\end{minipage}
\hspace{0.5mm}
\begin{minipage}[b]{0.3\textwidth}
 \includegraphics[width=\textwidth\centering]{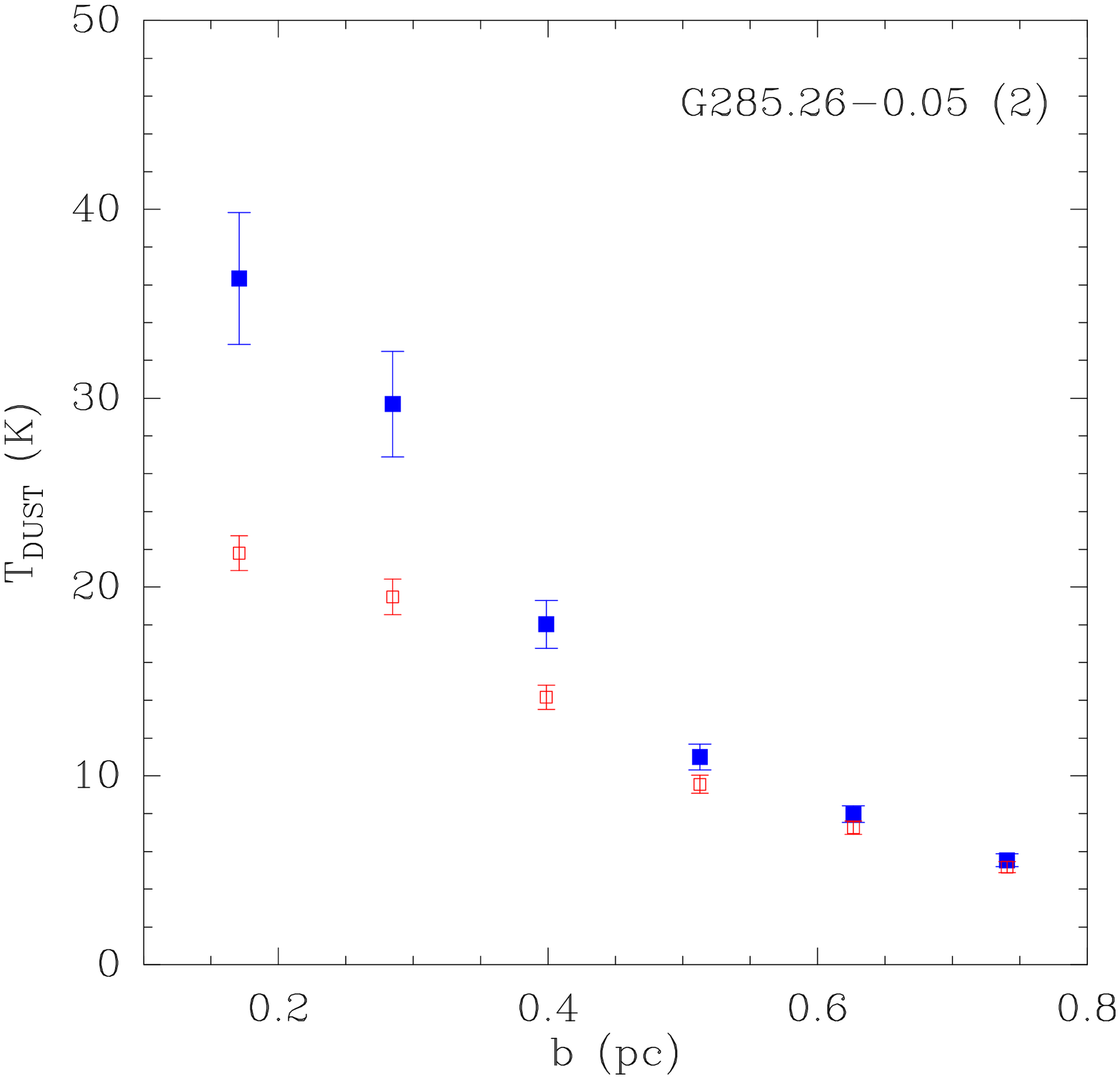}
\end{minipage}
\hspace{0.5mm}
\begin{minipage}[b]{0.3\textwidth}
 \includegraphics[width=\textwidth\centering]{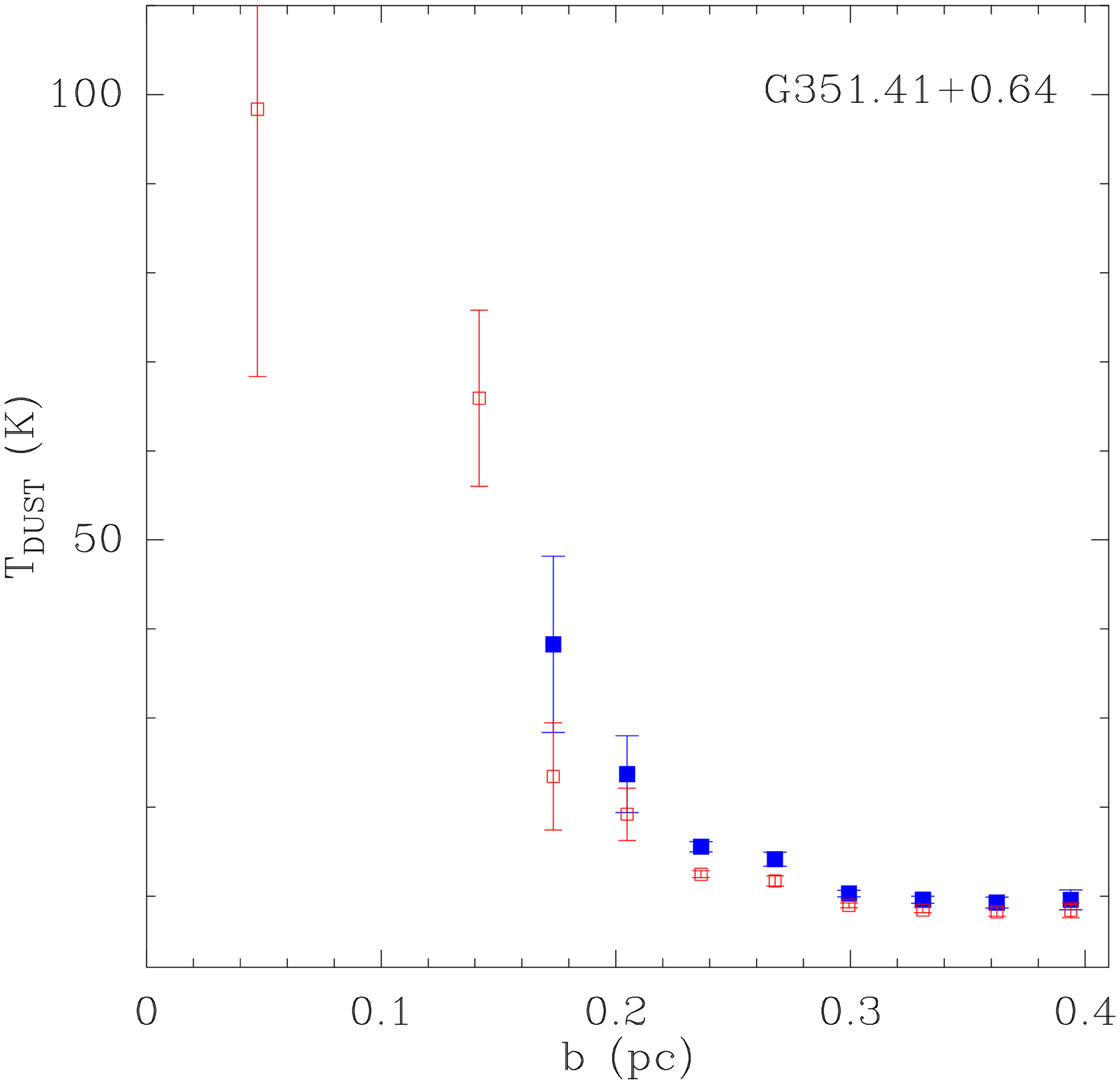}
\end{minipage}

\caption{
\footnotesize
Dust temperature profiles in the cores.
The filled blue and empty red squares correspond to $\beta=1.7$ and 2, respectively.
}
\label{1d_temp}
\end{figure*}

Figure~\ref{1d_temp} shows the dust temperature versus the impact parameter $b$
(distance from the emission peaks on the 350~$\mu$m maps) for $\beta=1.7$ and 2.
The coordinates of the 350~$\mu$m peaks
are indicated on the maps (see Fig.~\ref{tdust}).
The temperatures were averaged over concentric rings of the 5$''$ width taking into account individual errors
determined from the spread of the values relative to the average (see, e.g., \cite{Pir03} for a similar procedure description).
For G285.26 and G294.97, the dependencies are constructed separately for each core.
For the G294.97(1) core which has two clumps (1n and 1s),
the radial temperature profiles were calculated for each of them.

\section{Discussion}
\label{discussion}

As can be seen in Fig.~\ref{1d_temp}, the dust temperatures decrease with increasing
the impact parameter from $\sim 30-70$~K (for $\beta=1.7$) near the center to $\sim 10$~K at the periphery.
Dust temperatures for $\beta=2$ are systematically lower than for $\beta=1.7$,
and this difference decreases with distance from the center.
The highest dust temperatures are obtained towards the center of G351.41.
In most cores, the dust temperature averaged over the line of sight (except the value near the center) 
decreases almost linearly with impact parameter up to $\sim 0.2-0.7$~pc.
At the periphery, the temperature tends to $\sim 10$~K, which is lower than the values calculated
from fitting the $Herschel$ data in the 70--500~$\mu$m wavelength range \cite{vialactea,Marsh15,Marsh17} 
in the vicinity of the cores from our sample ($\sim 15-20$~K for $\beta=2$).
This difference may be due to the underestimation of our estimates because of the possible
systematic errors caused by the errors in absolute calibration, which, according to moderate estimates
(based on a comparison of the obtained fluxes from calibration sources) can reach $\sim 30$\%.
The presence of dust temperature variations on the line of sight
can be an alternative explanation.
In this case, $Herschel$ is sensitive to the dust emission with higher temperatures.
But even if dust temperatures are calculated by fitting the values in different ranges,
an accounting for the data at submillimeter and millimeter wavelengths, where the intensities are 2--3 orders
of magnitude less than the values in the far infrared range, will have little effect
on the resulting temperatures.
To verify this, the modified blackbody function was fitted into the flux values
for the G285.26(1) core at three wavelengths, namely, 350~$\mu$m, 500~$\mu$m, and 1.2~mm 
(355 Jy, 107 Jy, and 7 Jy, respectively).
The value at 1.2~mm was taken from \cite{Pir07}.
The value at 500~$\mu$m was calculated for the same area 
from the $Herschel$ data \cite{Herschel} using the SAOImageds9 package \cite{SAOImageds9}.
The temperature calculated from the least-squares fitting turned out to be 31~K
for $\beta=2$, underestimating the flux at 1.2~mm by $\sim 30$\%,
whereas the temperatures calculated from the flux ratios at 1.2~mm and 350~$\mu$m
and at 1.2~mm and 500~$\mu$m, from Eq.~(\ref{eq1})
for $\beta=2$ yields the values 18~K and 13.5~K, respectively.

The profiles of the dust temperature averaged over the line of sight,
obtained from observations (Fig.~\ref{1d_temp}), do not depend on possible systematic errors.
Let us consider whether they can be a result of the heating of dust by internal sources,
which exist in the studied cores (see Sec.~\ref{objects}).

\begin{figure}[t!]
\begin{minipage}[b]{0.42\textwidth}
 \includegraphics[width=\textwidth\centering]{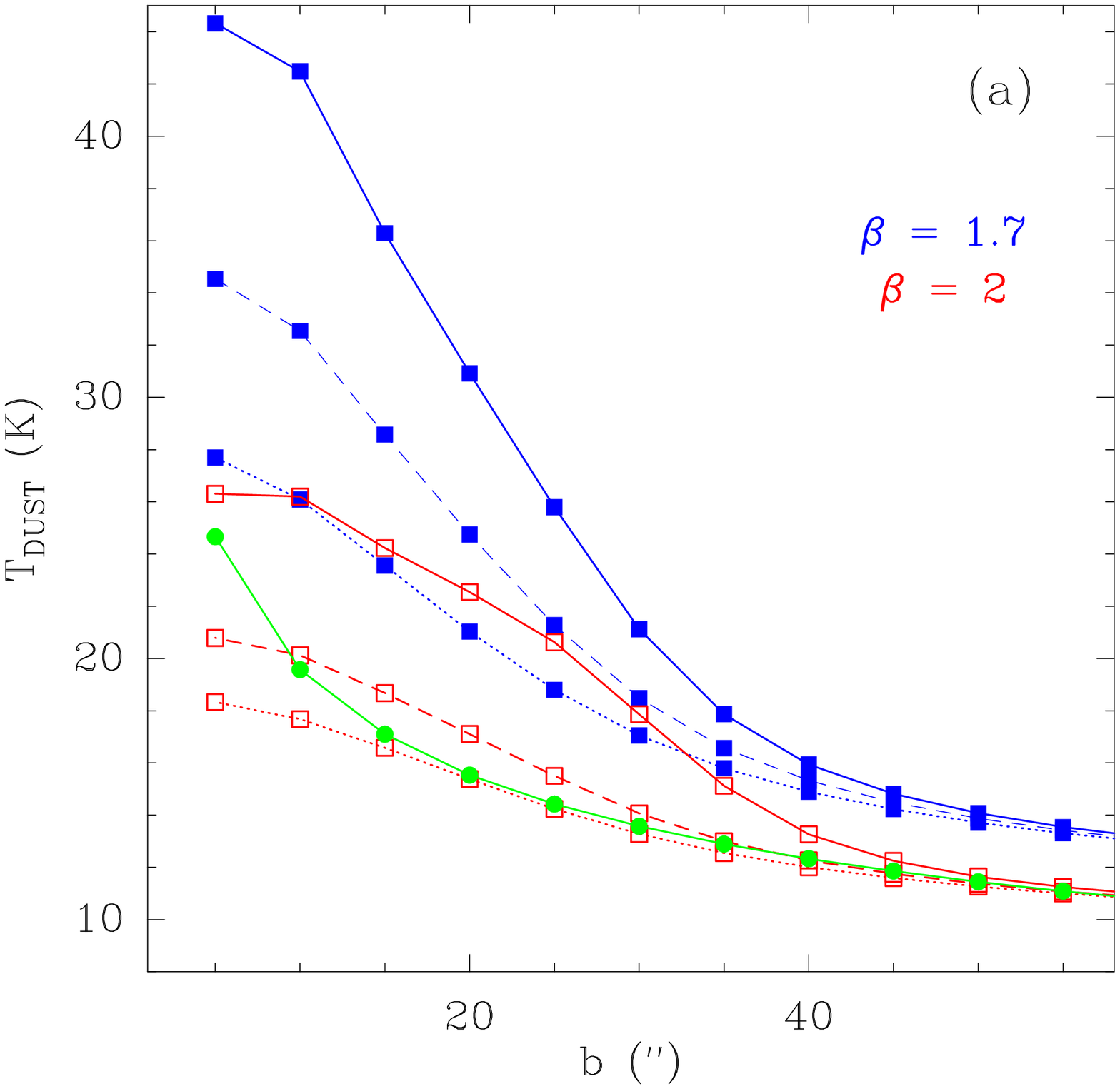}
\end{minipage}
\hspace{0.5mm}
\begin{minipage}[b]{0.45\textwidth}
 \includegraphics[width=\textwidth\centering]{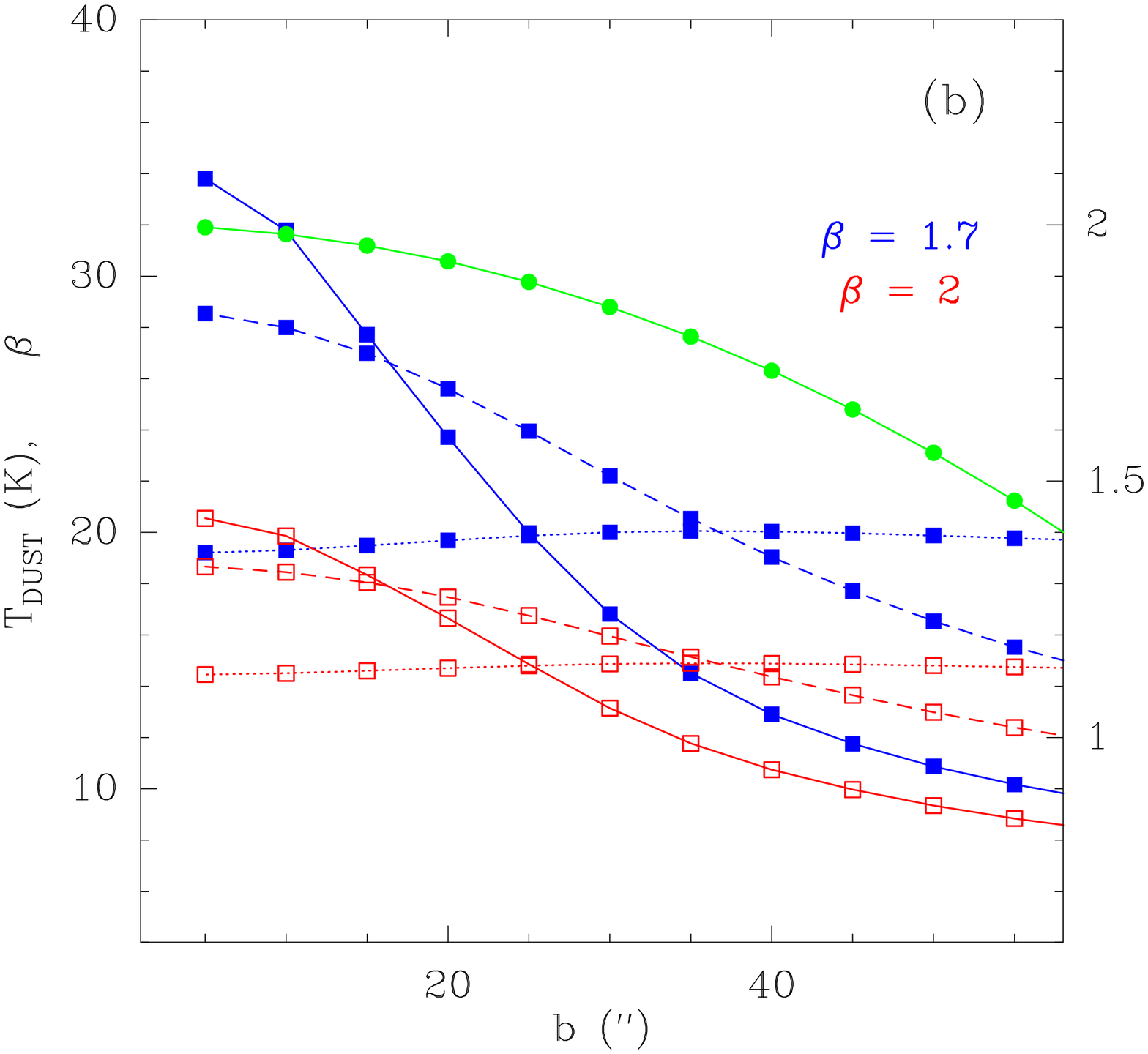}
\end{minipage}
\caption{
\footnotesize
Model dependencies of the dust temperature averaged over the line of sight on the impact parameter
calculated from Eq.~(\ref{eq1}) for two values of the $\beta$ index.
Panel $(a)$: the given radial dust temperature profile for $\beta_0=2$ is shown by green dots.
The dotted, dashed and solid curves correspond to the averaged dust temperature profiles
for the power-law index of the density profile $\alpha=1.5$, 2, and 2.5, respectively.
Panel $(b)$: the given radial profile of the $\beta$ index decreasing quadratically with increasing
distance from the center, is shown by green dots. 
The dotted curve corresponds
to the calculated dust temperature profile if the following dependence between
$\beta$ and temperature exist: $\beta=7\times T^{-0.47}$ \cite{Mannfors21}.
The dashed curve corresponds to a constant temperature of 20~K.
The solid curve corresponds to the given radial dust temperature profile: $\propto r^{-1/3}$.
The $\alpha$ index is equal to 2 in all cases.
}
\label{model}
\end{figure}

In the presence of an internal point source, the radial dust temperature profile in the core will have the form:
$\propto r^{-2/(4+\beta)}$ \cite{DL94}.
The dependence of the dust temperature averaged over the line of sight on the impact parameter
will also be nonlinear, but its specific form will depend on the type
of the density profile and the value of the telescope beam over which
the averaging is carried out.
To illustrate how various factors affect the observed profiles, calculations
were carried out within the framework of a simple spherically-symmetric model.
The intensity of dust emission for a given value of impact parameter
was calculated by integrating
over the line of sight the product of the reduced modified blackbody function and the density
radial profile ($\propto r^{-\alpha}$) at two wavelengths, 350~$\mu$m and 1.2~mm.
Then the obtained values were convolved with a two-dimensional Gaussian having a width of 24$''$
at the half maximum level.
Using Eq.~(\ref{eq1}), the averaged over the line of sight dust temperature was determined 
from the ratios of the intensities at two wavelengths for $\beta=1.7$ and 2,
depending on the impact parameter.
If the dust temperature in the core and the $\beta$ index are constant ($T_0$ and $\beta_0$, respectively),
then the temperature $T_{\rm DUST}$ calculated from Eq.~(\ref{eq1}) will also be constant,
coinciding with $T_0$ at $\beta=\beta_0$, and exceeding $T_0$ at $\beta<\beta_0$
regardless of the radial density profile.
If the dust temperature in the core depends on the radial distance
as $\propto r^{-2/(4+\beta_0)}$, then
the average temperature $T_{\rm DUST}$ will decrease with increasing impact parameter.
Figure~\ref{model}~(a) shows the calculated profiles of the dust temperature averaged
over the line of sight for $\beta=1.7$ and 2, and for $\alpha=1.5$, 2, and 2.5.
Due to the convolution with the beam, the nonlinear part
on the calculated profiles in the inner core regions
becomes close to the linear one (except for the region near the center), approaching the given
temperature profile on the periphery for $\beta=\beta_0=2$.
For $\beta<\beta_0$, the calculated temperature profiles lie higher than for $\beta=\beta_0$.
When approaching the center, the discrepancy between the calculated and the given temperature
profile increases, and the more the higher is the value of the power-law index of the radial density profile.
The higher the $\alpha$ value, the higher lie the $T_{\rm DUST}$ profiles
and the more nonlinear becomes the calculated profile.
Thus, the observed profiles of the dust temperature averaged over the line of sight
in the studied cores can, in principle, be explained by the heating from an internal source
selecting appropriate values of the parameters $\beta$ and $\alpha$.

An alternative hypothesis for explaining the observed profiles may be the existence of variations
of the $\beta$ index in the cores due, e.g., to possible differences in the effective size
and composition of dust particles at the center and at the periphery, as well as the presence
of an inversely proportional dependence between $\beta$ and dust temperature.
In this case, the $T_{\rm DUST}$ profiles calculated from Eq.~(\ref{eq1}) under the assumption
of a constant $\beta$ value, can differ significantly from the given temperature profile.
Calculations performed within the framework of the spherically-symmetric model
with a radial profile of $\beta$ give the following results.
In the absence of a heating source and assuming $\beta$ to decrease with distance
from the center according to some arbitrary chosen power law (Fig.~\ref{model}(b), green points),
as well as in the presence of inversely proportional dependence between $\beta$ and dust temperature
of the form: $\beta=7\times T^{-0.47}$ (this dependence was obtained in \cite{Mannfors21}
from the analysis of observational data of several hundreds of the $Planck$ dense clumps),
the temperatures averaged over the line of sight almost do not depend on the impact parameter
(Fig.~\ref{model}(b), dotted curves), since different gradients of temperature and $\beta$
have a mutually compensating effect on the $T_{\rm DUST}$ profiles.
Replacing the $\beta$ gradient from descending to ascending, or changing the density profile
while keeping the same dependence between $\beta$ and temperature, almost does not influence
the calculated profiles.
If one assumes that the dust temperature and $\beta$ do not depend on each other,
while $\beta$ still decreases to the edge and the temperature is constant,
then the calculated $T_{\rm DUST}$ profile will correlate with the given $\beta$ radial profile,
decreasing with distance from the center (Fig.~\ref{model}(b), dashed curves).
In the presence of a temperature profile determined by heating from an internal source,
the drop in $T_{\rm DUST}$ with impact parameter will be sharper
(Fig.~\ref{model}(b), solid curves)
or weaker if $\beta$ is increasing to the edge.

The analysis shows that the drop in the dust temperature averaged over
the line of sight, derived from the ratios of intensities at two wavelengths,
with increasing of impact parameter is due, most probably, to the presence
of an internal source in the cores.
Although the specific type of the calculated temperature profiles depends on
the accepted value of the $\beta$ index, as well as on the density profile in the core
and the size of the telescope main beam, their general behavior does not change.
At the same time, the higher the $\alpha$ value, the higher lie the calculated profiles.
In the G351.41 core, where the calculated temperature profile is to a greater extent nonlinear
compared to other cores (Fig.~\ref{1d_temp}), the power-law index
of the density profile can be quite high ($\alpha\ga 2.5$).
If one assumes that the $\beta$ index in the studied cores is not constant
and does not depend on temperature, the decline of the calculated temperature
averaged over the line of sight with impact parameter
can occur if $\beta$ decreases from the center to the edge,
while the temperature in the core is either constant or decreases to the edge.
The existence of such dependence of $\beta$ on radial distance
may be due to gradients in the size (and composition) of dust particles
inside the cores, which, in turn, can be caused by the processes of suppression
of the growth of dust grains, evaporation of their mantles,
or destruction of large grains due to the increased level of turbulence
in the inner regions of the cores \cite{Tang21,Hirashita09}.
However, this also requires the existence of an internal source.

Thus, although the dust temperature estimates derived from observational data
at two wavelengths depend on the accepted value of $\beta$ and can differ greatly
from the real ones, spatial distributions of the intensity ratios can be used
to study the nature of the dust temperature variation in dense cores.
An existence of compact regions with enhanced dust temperatures can serve
an indicator of the early stages of the star formation process,
and this fact can be used when selecting objects for detailed research.
To answer the question of how the $\beta$ index, determined by physical characteristics
and composition of dust, varies inside the cores associated with the regions of massive star
and star cluster formation, further observational studies in different wavelength ranges
with high angular resolution and sensitivity are necessary and a comparison with the results
of chemical and gas-dynamic modelling taking into account dust evolution
in the cores and various factors, including supersonic turbulence, is needed.

\section{Conclusions}

Using the APEX-12m telescope, we obtained maps in continuum at a wavelength of 350~$\mu$m
of eight gas-dust clouds from the southern hemisphere.
The clouds are associated with the regions of massive star and star cluster formation
and have dense cores.
The core sizes estimated at the half maximum intensity level at 350~$\mu$m
are $\sim 0.1-0.2$~pc, the masses of the cores and their mean gas densities
lie in the ranges: $\sim 20-1000~M_{\odot}$ and $\sim (0.3-7.3)\times 10^6$~cm$^{-3}$, respectively.
The 350~$\mu$m data were compared with the data at 1.2~mm \cite{Pir07}.
From the ratios of intensities at two wavelengths convolved to the same angular resolution,
the dust temperatures averaged over the line of sight were calculated and their spatial distributions were obtained.
Dust temperature maps in most cores correlate with intensity maps at 350~$\mu$m.
The temperatures at the core periphery ($\sim 10$~K) are lower than the estimates obtained from
the $Herschel$ data, which can be due to both systematic errors in our observations
and variations of dust temperature on the line of sight.
A decrease of the dust temperature averaged
over the line of sight with a distance from the center is found.
The obtained dust temperature profiles are close to linear ones in most cases.
Using a simple spherically-symmetrical model, it is shown that the estimates
of the temperatures averaged over the line of sight stronly depend on the chosen value $\beta$,
the index of the power-law dependence of the dust emissivity on frequency, as well as
on the type of the temperature and density radial profiles.
Temperature profiles similar to the observed ones can be obtained in a model
with internal source by varying the parameters of the density profile
and taking into account convolution with the telescope beam.
If $\beta$ varies with a distance from the center which is related
with variations of size and composition of dust particles,
then the dust temperature profiles, similar to the observed ones,
can be obtained in the case of the $\beta$ decline from the center to the edge,
which, in turn, may be caused by the presence of an internal source.
Thus, the presence of an internal source appears to be the main reason
for the observed temperature profiles in the studied cores.

\section{Acknowledgements}

The author is grateful to the reviewers for valuable questions and comments,
which led to conducting an additional search for inner sources in the sample objects,
carrying out new model calculations, and to significant revision of the sections
with description of the objects and discussion of the results.
The author thanks Attila Kov\'acs (Caltech) for consulting on the miniCRUSH program
and kindly providing an updated version of the program,
as well as Peter Zemlyanukha for the help with installation of the BoA program.
The processing and analysis of observational data was carried out as a part
of the implementation of the State assignment for the IAP RAS (project No. 0030-2021-0005).

{}


\end{document}